\documentclass[twocolumn,astrosymb,tighten,longbib,raggedbottom]{aastex7} 
\makeatletter
\g@addto@macro\normalsize{%
  \setlength\abovedisplayskip{5pt plus 2pt minus 3pt}%
  \setlength\belowdisplayskip{5pt plus 2pt minus 3pt}%
  \setlength\abovedisplayshortskip{0pt plus 2pt}%
  \setlength\belowdisplayshortskip{3pt plus 2pt minus 2pt}}
\makeatother
\usepackage{mathrsfs}

\usepackage{nccmath,gensymb} 
\usepackage{siunitx, xspace,tabularx,multirow}

\allowdisplaybreaks

\newcommand{\um}{\,\micron\xspace}
\newcommand{\healpix}{HEALPix\xspace}
\makeatletter
\def\paragraph{\@startsection{paragraph}{4}{\z@}%
  {0.1ex}{0pt}{\small\itshape}}   
\makeatother

\shorttitle{SPHEREx PAH Forward Modeling}
\shortauthors{Saydjari et al.}

\graphicspath{{./}{figures/}}

\begin{document}

\title{Forward modeling linear variable filter data: \\[0.5ex] High resolution PAH and Paschen-$\alpha$ maps from SPHEREx}

\author[orcid=0000-0002-6561-9002,gname=Andrew,sname=Saydjari]{Andrew~K.~Saydjari}
\altaffiliation{Hubble Fellow}
\affiliation{Department of Astronomy, New Mexico State University, Las Cruces, NM 88001 USA}
\affiliation{Department of Astrophysical Sciences, Princeton University,
Princeton, NJ 08544 USA}
\email[show]{aksaydjari@gmail.com}  
\correspondingauthor{Andrew~K.~Saydjari}

\author[orcid=0000-0002-2506-2502,gname=Connor,sname=Hainje]{Connor~Hainje}
\affiliation{Center for Cosmology and Particle Physics, Department of Physics, New York University, 726 Broadway, New York, NY 10003, USA}
\email{connor.hainje@nyu.edu}

\author[orcid=0000-0002-2519-2219,gname=Benjamin,sname=Dodge]{Benjamin~Dodge}
\affiliation{Department of Physics, Stanford University, Stanford, CA 94305, USA}
\affiliation{Kavli Institute for Particle Astrophysics \& Cosmology, P.O. Box 2450, Stanford University, Stanford, CA 94305, USA}
\email{bendodge@stanford.edu}

\author[orcid=0000-0001-5610-3779,gname=Philipp,sname=Frank]{Philipp~Frank}
\affiliation{Max Planck Institute for Astrophysics, Karl-Schwarzschild-Str. 1, 85748 Garching, Germany}
\affiliation{Department of Physics, Stanford University, Stanford, CA 94305, USA}
\affiliation{Kavli Institute for Particle Astrophysics \& Cosmology, P.O. Box 2450, Stanford University, Stanford, CA 94305, USA}
\email{philipp@mpa-garching.mpg.de}

\author[orcid=0000-0003-2866-9403,gname=David,sname=Hogg]{David~W.~Hogg}
\affiliation{Center for Cosmology and Particle Physics, Department of Physics, New York University, 726 Broadway, New York, NY 10003, USA}
\affiliation{Max Planck Institute for Astronomy, K\"onigstuhl 17, Heidelberg 69117, Germany}
\affiliation{Center for Computational Astrophysics, Flatiron Institute, 162 Fifth Ave, New York, NY 10010, USA}
\email{david.hogg@nyu.edu}  

\author[orcid=0000-0002-7633-3376,gname=Susan,sname=Clark]{Susan~E.~Clark}
\affiliation{Department of Physics, Stanford University, Stanford, CA 94305, USA}
\affiliation{Kavli Institute for Particle Astrophysics \& Cosmology, P.O. Box 2450, Stanford University, Stanford, CA 94305, USA}
\email{susanclark@stanford.edu}

\begin{abstract}\noindent
Polycyclic aromatic hydrocarbon (PAH) emission is a powerful tool for the study of the interstellar medium (ISM) in both Galactic and extragalactic contexts.
It can reveal ionization state, radiation field, metallicity, composition, and ISM structure as sculpted by processes like turbulence and stellar feedback. Here we use NASA's SPHEREx to produce clean, high-resolution PAH maps at spectral resolution R $\sim$ 35 (compared to WISE at R $\sim$ 2). Prior treatments of the SPHEREx linear variable filter data seek to convert them to more conventional band-averaged maps that can leave striping artifacts associated with spatially variable wavelength sampling, especially at emission lines. In contrast, we deliver a scalable forward-modeling method capable of generating high-quality all-sky SPHEREx emission-line maps at $8\farcs5$ FWHM angular resolution (on $6\farcs4$ pixels), while naturally producing maps without visible striping. We reformat the data into a pixel-level database, mask stellar contamination, and infill missing pixels with a conditional, data-driven, anisotropic Gaussian process. We apply our method to produce 3.3\um and 3.4\um PAH and Paschen-$\alpha$ emission maps of a set of fields spanning a range of Galactic and extragalactic conditions including the Rosette nebula, the Corona Australis molecular cloud, M33, and M31. We demonstrate the quality and synergies of our SPHEREx 3.3\um PAH maps by comparisons with other available gas and dust tracer maps and achieve sensitivities up to $\sigma(I_{3.3}) = 2.7 \times 10^{-8}$ erg s$^{-1}$ cm$^{-2}$ sr$^{-1}$. We estimate that a two-year extension of the SPHEREx survey, with modest modifications to the scanning strategy, would match or exceed the PAH sensitivity of WISE over the whole sky.
\end{abstract}


\section{Introduction} \label{sec:intro}

\begin{figure*}[p]
    \centering
    \includegraphics[width=\linewidth]{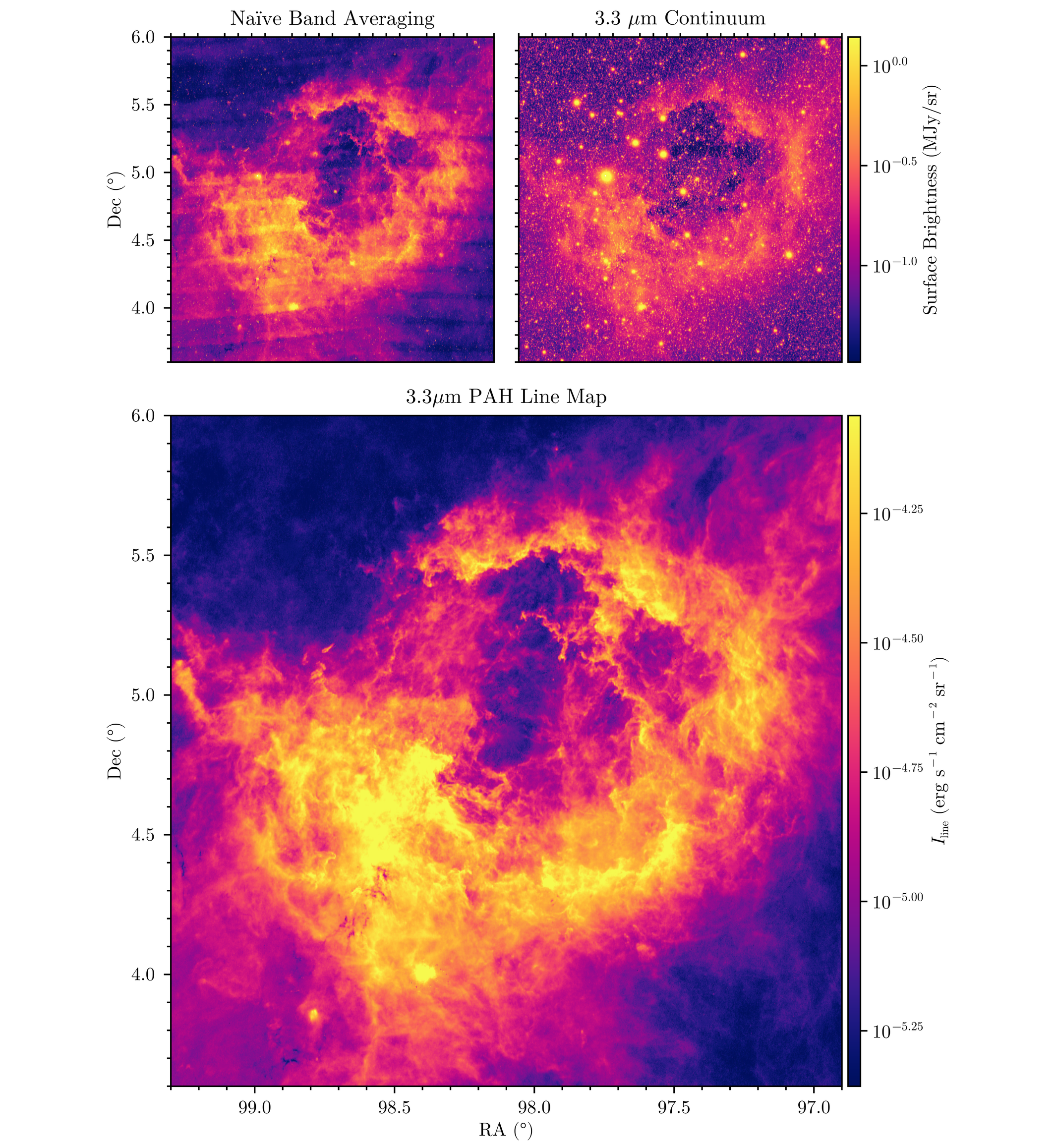}
    \caption{A demonstration of our forward-modeling approach applied to the Rosette nebula: (top right) surface brightness of our continuum model at 3.29\um and (bottom) the 3.3\um polycyclic aromatic hydrocarbon (PAH) line surface brightness. For comparison, a straightforward band average of observations near 3.3\um minus continuum bands straddling the line is shown in the top left.}
    \label{fig:rosette_methodcompare_full}
\end{figure*}

The interstellar medium (ISM) is a chemically and dynamically rich environment. Polycyclic aromatic hydrocarbons (PAHs) are a ubiquitous class of interstellar organics that play an outsized role in the energetics of the ISM \citep{Tielens_2008_ARAA}. PAHs are stochastically heated by the interstellar radiation field, and produce prominent ro-vibrational emission lines in the mid-infrared that are widely used as tracers of interstellar physics, both in the Milky Way and extragalactically \citep{Baron_2025_ApJ}. In the Milky Way, diffuse PAH emission is an exquisite tracer of high-angular resolution structure, enabling constraints on interstellar turbulence \citep{Miville-Deschenes_2016_AA, Vigoureux:2026}, the relative orientation of density and magnetic field structures on small scales \citep{CordovaRosado_2024_ApJ}, and other physics that is encoded in the small-scale spatial structure of the ISM. These analyses require separating the diffuse PAH emission from other near-infrared emission, including from stellar and extragalactic sources. \citet{Meisner_2014_ApJ} produced 12\um diffuse emission maps from Wide-field Infrared Survey Explorer (WISE; \citealt{Wright_2010_AJ}) data that have been widely used as tracers of diffuse PAH emission. In nearby galaxies, spatially resolved maps of PAH emission can be used alone or in conjunction with other tracers to study the ISM properties, star formation, and the radiation field across galactic environments.  Table~\ref{tab:pah_features} shows a selection of PAH features and the types of molecules predominantly responsible for the observed emission.

\begin{table}[ht]
  \centering
  \caption{Relevant PAH emission features compiled from
  \citet{Draine_Li_2007} and \citet{Tielens_2008_ARAA}. Charge and size refer to the
  molecules predominantly responsible for the observed emission and are
  not a property of the features themselves. Our work maps the 3.3 and
  3.4~$\mu$m features, but the other features are relevant for
  comparisons with maps from WISE and the Spitzer Infrared Array Camera \citep[IRAC;][]{Fazio_2004_ApJS}
  (see Section~\ref{sec:galactic}).}
  \label{tab:pah_features}
  \footnotesize
  \begin{tabular*}{\columnwidth}{@{\extracolsep{\fill}} l l l l l @{}}
    \hline
    $\lambda$ ($\mu$m) & Mode & Bond type & Charge & Size \\
    \hline
    3.3  & C--H stretch   & Aromatic  & Neutral & Small  \\
    3.4  & C--H stretch   & Aliphatic & ---     & Small  \\
    6.2  & C--C stretch   & Aromatic  & Ionized & Medium \\
    7.7  & C--C stretch + & Aromatic  & Ionized & Medium \\
         & C--H bend      &           &         &        \\
    8.6  & C--H bend      & Aromatic  & Ionized & Medium \\
         & (in-plane)     &           &         &        \\
    11.3 & C--H bend      & Aromatic  & Neutral & Large  \\
         & (out-of-plane) &           &         &        \\
    12.7 & C--H bend      & Aromatic  & Neutral & Large  \\
         & (out-of-plane) &           &         &        \\
    \hline
  \end{tabular*}
\end{table}

The Spectro-Photometer for the History of the Universe, Epoch of Reionization and Ices Explorer (SPHEREx) is a NASA Explorer satellite that is currently conducting an all-sky, near-infrared (0.75 to 5.0\um) spectral survey \citep{Bock_2026_ApJ}. These data present rich opportunities for ISM science: the SPHEREx spectral range encompasses the bright 3.3\um aromatic PAH feature, as well as the 3.4\um aliphatic hydrocarbon feature, hydrogen recombination lines, several ro-vibrational transitions of molecular hydrogen, and ice absorption features. In this work, we develop a method for forward-modeling the SPHEREx data that enables high-fidelity mapping of diffuse-emission spectral features, and apply our method to construct high-resolution maps of the 3.3\um and 3.4\um PAH features and Paschen-$\alpha$. For simplicity, we refer to this method as LVFit throughout the paper. 

\begin{figure*}[p]
    \centering
    \includegraphics[width=\linewidth]{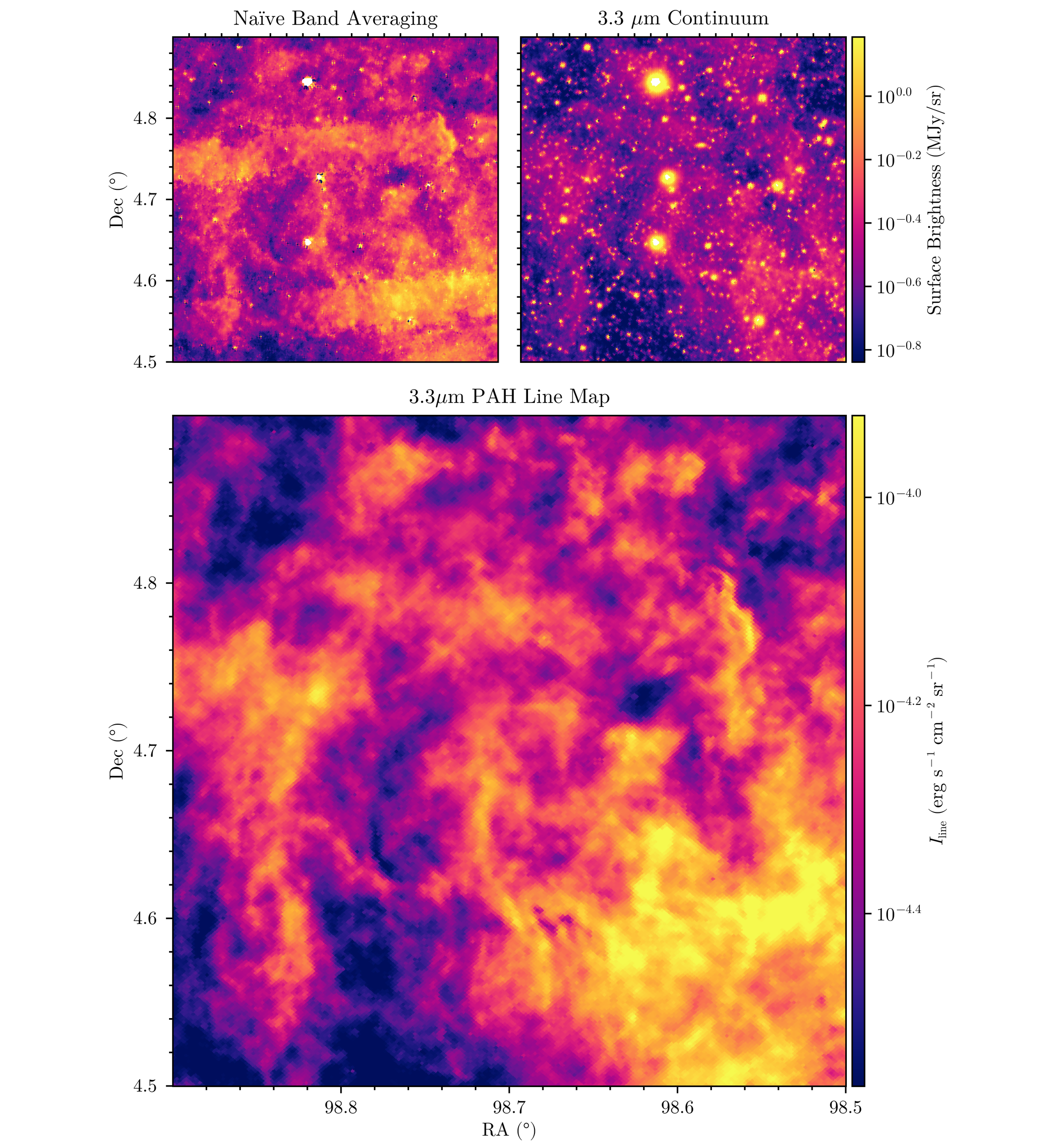}
    \caption{Same as Figure~\ref{fig:rosette_methodcompare_full}, but zoomed in to the middle left.}
    \label{fig:rosette_methodcompare_small}
\end{figure*}

SPHEREx uses linear variable filter (LVF) spectrometers, which gradually vary the spectral transmission across the surface of the detectors. This means that each pixel in the resulting spectral image corresponds to a slightly different wavelength and bandwidth. The SPHEREx team defines 17 bands for each of the 6 LVFs, which divides the data into 102 spectral channels \citep{Bock_2026_ApJ, SPHEREx_ExpSupp_2026}. Much of the work on tracing PAH emission with SPHEREx to date has operated in this paradigm, by binning the measurements into spectral channel maps (see Section~\ref{sec:compare2others}). Naïve binning of the data can give rise to banding artifacts as shown in Figure~\ref{fig:rosette_methodcompare_full} (zoomed in Figure~\ref{fig:rosette_methodcompare_small}), which are also prominently visible in the low-resolution SPHEREx PAH maps produced by \citet{Cepeda-Arroita_2026_arXiv} (see Section~\ref{sec:compare2others}). 

Here we produce large-area maps from many individual exposures using a forward-modeling method. Standard methods in optical astronomy for this kind of ``mosaicing'' or map-making traditionally involve interpolation and median (or similar) averaging of overlapping imaging; the current state of the art is drizzle \citep{drizzle}. These approaches are inappropriate here because the LVFs make it such that every pixel of the detector has (in principle) a slightly different spectral response. If images are made by binning pixels into spectral channels and combining by traditional methods, banding appears. Since the readout from any pixel of any detector in any good astronomical telescope is linearly related to the surface brightness on the sky, it is principled to make maps of the surface brightness by a classical ``inverse problem'' \citep{inverse} or forward model, from which the map is found by optimizing an objective function based on a justified likelihood function.

These forward-model map-making methods have excellent properties: The output maps are linearly (or nearly linearly, depending on details) related to the input data, such that measurements made in the map are simply related to measurements made in the original raw data. The output maps can saturate the Cram\'er--Rao bound \citep{rao45, cramer46} on the delivered image information. The input data can be arbitrarily heterogeneous in spatial structure, noise properties, and spectral response, provided that the likelihood function or forward model accounts for these properties correctly. Forward-model methods for map-making have had an extremely high impact in cosmic microwave background experiments (notably pioneered by \citealt{wright96}, and used in different forms by the Wilkinson Microwave Anisotropy Probe (WMAP; \citealt{wmapmaps}), Planck (\citealt{planckmaps}), and the Atacama Cosmology Telescope (ACT; \citealt{actmaps})). Recently these ideas have been introduced into optical imaging and spectroscopy \citep{tractor, Saydjari_2023_ApJ, frizzle}.

One challenge or question with forward-modeling approaches is:
Do we choose to deconvolve (in any sense) when we combine the raw data, or do we create a map at the native resolution of the raw data?
Since a forward model is in play, the parameters of that model can be of the surface brightness field at any resolution (within reason), in either the angular or spectral sense. When the data and model are both good, the inverse problem can deliver a map at higher resolution than the raw data, at the cost of dependency on detailed assumptions.
Here we use beliefs about the PAH and recombination lines to effectively perform only a light deconvolution of the SPHEREx data in the spectral direction. That is, we measure the PAH and recombination lines as precisely as possible given what we know about those line positions and shapes, and about the spectral response of each pixel of the SPHEREx hardware.

The rest of this paper is structured as follows. In Section~\ref{sec:datamanage} we describe the SPHEREx data products used in this work, and our processing of these data to prepare them for use in our forward model. We describe our methodology in detail in Section~\ref{sec:methods}. We provide a detailed look at maps of select Galactic regions (Section~\ref{sec:galactic}: the Rosette nebula, Corona Australis, Cygnus X, and the Heart and Soul nebulae) and external galaxies (Section~\ref{sec:extragalactic}: M31 and M33). We measure the sensitivity of our map and compare to other literature SPHEREx maps in Section~\ref{sec:disc}. We discuss the limitations of our methodology and future improvements in Section~\ref{sec:conc} and make these maps publicly available in Section~\ref{sec:dataavil}. 

\section{Data Management} \label{sec:datamanage}

This work makes use of the SPHEREx Quick Release 2 (QR2) and Quick Release 3 (QR3) data products. These data are the Level 2 products of the SPHEREx data processing pipeline \citep{Akeson_2026_ApJS}, which performs astrometry, flux calibration, and bad pixel flagging. The resulting data are ``spectral images'', consisting of calibrated surface brightness measurements, variance and zodiacal light (zodi) estimates, and status flags.

The dataset used in this work includes all exposures through 2026 week 36 (2025 April 24 to 2026 September 7). QR2 consists of all exposures taken up to 2026 week 29; QR3 is all data taken starting 2026 week 30 \citep{SPHEREx_ExpSupp_2026}. The two release versions differ by a major upgrade to the reduction pipeline, which updated the calibrations, applied a new data-driven rescaling to zodiacal light estimations, and enhanced the bad-pixel flagging. The updated calibrations and zodiacal light rescalings are accompanied by calibration products (absolute gain matrix, dark current map, flux correction, zodi scaling) that allow retroactively updating the QR2 data.
For a measurement from detector pixel $j$ with surface brightness $F$, variance $V$, and zodi $Z$, the QR3-recalibrated measurements are given by
\begin{equation} \label{eq:qr3recal}
\begin{aligned}
    F_{\rm QR3} &= c_j \, F_{\rm QR2} + A_{{\rm QR3},j} \, (D_{{\rm QR2},j} - D_{{\rm QR3},j}), \\
    V_{\rm QR3} &= c_j^2 \, V_{\rm QR2}, \\
    Z_{\rm QR3} &= \alpha(\lambda_j) \, Z_{\rm QR2},
\end{aligned}
\end{equation}
where
$c$ is the flux correction (which rescales the gains from QR2 to those of QR3),
$A$ is the absolute gain matrix,
$D$ is the dark current map,
and $\alpha$ is the zodi scaling, linearly interpolated to the central wavelength $\lambda_j$ of pixel $j$.

Because of the LVF design of SPHEREx, a spectral image mixes spatial and spectral information. In order to more directly query the data by sky position, we pre-process all QR2 and QR3 data into a pixel-level database called the \texttt{talltable}, the details of which are in preparation. To summarize, the pixels of every image are flattened into a set of tuples containing the RA, Dec, surface brightness, variance, flags, zodi, and detector position (detector number, pixel row, pixel column) of each pixel. We call these tuples ``pixel measurements.'' We then compute the NESTED, $N_{\rm side} = 2^{22}$ \healpix index \citep{Gorski_2005_ApJ} of the sky position and sort the pixel measurements from all images according to this index.

There are in total $6.3\times10^{12}$ pixel measurements in our dataset, so the data are split into chunks called ``partitions''. Partitions are defined by coarser \healpix indices: $N_{\rm side} = 2^{6}$ over most of the sky, with adaptive refinement up to $N_{\rm side} = 2^{9}$ in the two deep fields at the ecliptic poles, designed to keep the number of measurements per partition under 200 million. In each partition we store the pixel measurements in Parquet files \citep{Apache_Parquet}, which enable very fast queries and filters. Data that apply only to individual images (such as image metadata) or to individual detector positions (such as the wavelength solution) are stored in separate Parquet files. Each pixel measurement carries an ID for its image and detector position, so these data can be joined with the pixel measurements at query time. 

\begin{figure*}[t]
    \centering
    \includegraphics[width=\textwidth]{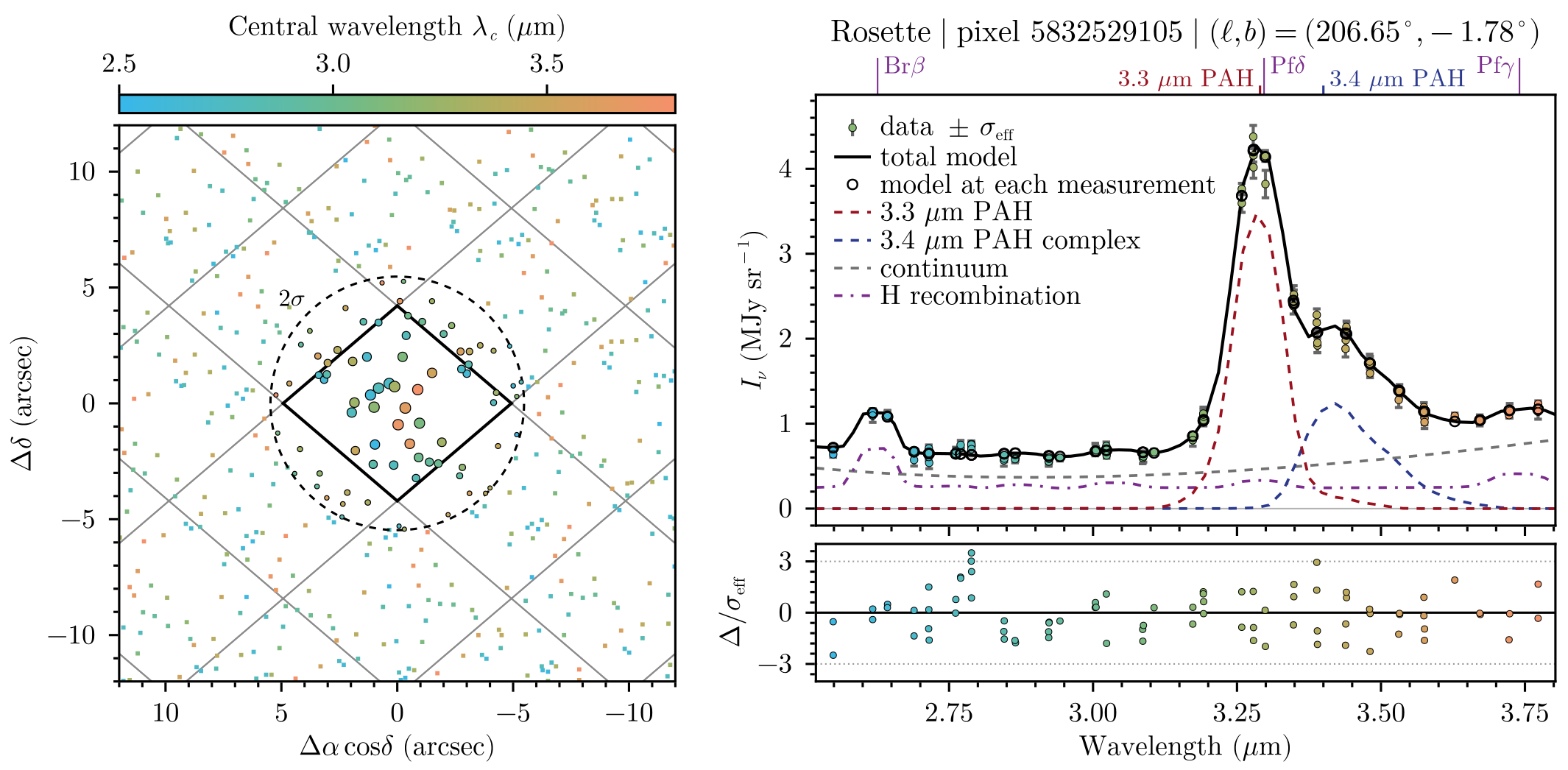}
    \caption{The LVFit model of one map pixel in the Rosette nebula (Index 5832529105, \healpix $N_{\rm side} = 2^{15}$, NESTED). Left: The pixel of interest (black) and its neighbors (gray), with the $2\sigma$ limit circle used to decide on including measurements. Every SPHEREx measurement within the field of view that passes our wavelength cut for the PAH model is shown, colored by its central wavelength. Measurements inside the circle that enter the fit are drawn as circles, with areas that scale with their spatial weight. Measurements not included in the fit are drawn as squares of one size. Right: The same 80 measurements in spectral space with their effective uncertainties. We show the total model (black) and its components as the LVF would see them. Open circles mark the model evaluated with the response of each individual measurement. The positions of the PAH features and of the three strongest hydrogen recombination lines in the window are marked above the panel. Pf$\delta$ falls on the 3.3\um feature. The lower panel shows the normalized residuals. For only two of the 80 measurements do the effective uncertainties include a down-weighting due to the iteratively reweighted least squares procedure of Section~\ref{sec:IRLS}.}
    \label{fig:pixel_schematic}
\end{figure*}

\section{Methods} \label{sec:methods}

\subsection{Linear Fit} \label{sec:linfit}

At its core, our map-making process LVFit is a least-squares linear fit of a set of basis vectors to all of the SPHEREx observations at a given location on the sky. We make a map by measuring the amplitude of each basis vector on a grid of \healpix $N_{\rm side} = 2^{15}$ ($\sim6\farcs4$) pixels, which have an area of $41.5$ arcsec$^2$. For each pixel, we gather all pixel measurements from the \texttt{talltable} (described in Section~\ref{sec:datamanage}) that lie within $5\farcs5$ great-circle distance of the pixel center (Figure~\ref{fig:pixel_schematic}). We also select only observations with wavelengths within a model-dependent window around the features of interest, measured in units of each measurement's own bandwidth (see Section~\ref{sec:models}), and we require measurements to have a finite surface brightness, positive variance, and no bit of the SPHEREx quality flag \citep{SPHEREx_ExpSupp_2026} set other than \texttt{SOURCE} (bit 21), which only marks pixels mapped to a known catalog source. We keep those observations because stars are handled by our own mask (Section~\ref{sec:mask}). 

We additionally give zero weight to the measurements of the detector pixels in the SPHEREx non-functional pixel list \citep{SPHEREx_ExpSupp_2026}, in our own list of pixels with anomalous residuals (released in Section~\ref{sec:dataavil}), and to the measurements rejected by the contamination tests of Section~\ref{sec:contam}. We subtract the Level 2 zodiacal light estimate from every measurement (Section~\ref{sec:datamanage}). We weight the observations as a function of their great-circle distance from the center of the pixel being fit, using a Gaussian with a FWHM of one pixel, $6\farcs44$ ($\sigma = 2\farcs73$), truncated at $2\sigma = 5\farcs5$. Adding the $\sim6\arcsec$ pixel-integrated point spread function (PSF) of SPHEREx \citep{Bock_2026_ApJ} in quadrature to this spatial weighting predicts an angular resolution of $\sim9\arcsec$ FWHM. This agrees with the $8\farcs5$ FWHM we measure empirically for isolated stars in the LVFit continuum maps, which we quote as the angular resolution of the maps. 

Figure~\ref{fig:pixel_schematic} illustrates this selection and the resulting fit for one pixel in the Rosette nebula. This weight is multiplied by the reported inverse variance to form $w_i^{\rm data}$ for each measurement $i$. The amplitude of each basis vector is then obtained in a first pass simply by weighted least squares, which is refined as described in Section~\ref{sec:IRLS}. Our basis vectors are defined as functions of wavelength (on a 2.5~nm grid), but are compared to the data after a transformation given by the spectral response model of the SPHEREx LVF (see Section~\ref{sec:LVF}). In adopting this procedure, we are making several simplifying assumptions that are not true in reality. We discuss these assumptions, their limitations, and possible solutions in Section~\ref{sec:conc}.

It is important to note that the spatially varying banding we describe in Section~\ref{sec:intro} that is inherent to naïve binning of the data can also appear in a linear forward model, and will if the model is misspecified. If either the basis vectors or spectral response cause the model to poorly describe the data, then the phase of the wavelength sampling relative to the spectral features of interest will bias the recovered amplitudes in a spatially varying manner. As a concrete example, if the intrinsic width of a basis vector peak is underestimated, then the amplitude you would infer for that peak will depend on whether the wavelength samples you observe straddle either side of the maximum or if a measurement falls at the wavelength center of the peak. This is relevant for the SPHEREx survey strategy because within one survey pass, the dithers and consecutive pointings step a given sky position through the LVF in a regular ladder of wavelengths whose rungs are a fixed 2.7\% apart in wavelength, close to one resolution element ($R \approx 35$), because the LVF wavelength is nearly exponential in detector position \citep{Bock_2026_ApJ, SPHEREx_ExpSupp_2026}. Except in the deep ecliptic fields, each later pass lays down a new ladder at a different phase rather than repeating the earlier wavelengths. The phases of those ladders relative to a feature therefore vary smoothly across a field, and any misspecification of the feature shape turns into smooth, spatially varying stripes.

\subsubsection{Iteratively Reweighted Least Squares} \label{sec:IRLS}
In order to better handle outliers, for example from unmasked bad pixels, transient unmasked cosmic rays, or stray PSF wings, we solve the linear problem above using iteratively reweighted least squares \citep[IRLS;][]{Green_1984_JRSSB, Holland_1977_CommStat}. This involves solving the linear problem, updating the weights to down-weight measurements that have large residuals, and repeating until some convergence criterion is met or a maximum number of iterations is reached.

\begin{equation} \label{eq:irwls}
\begin{aligned}
    w_i^{\rm robust} &= \min\!\left(1,\ \frac{k}{\sqrt{w_i^{\rm data}}\,|r_i|}\right), \\
    w_i^{\rm new} &= w_i^{\rm robust}\, w_i^{\rm data}
\end{aligned}
\end{equation}

Here $w_i^{\rm data}$ is the original data weight for measurement $i$, in our case the combined spatial and inverse variance weight, $r_i$ are the residuals, and $w_i^{\rm new}$ is the weight to be used in the next round of least squares fitting. When the normalized residual $\sqrt{w_i^{\rm data}}\,|r_i|$ is less than $k$, the weight is unchanged. When it is greater than $k$, the measurement is down-weighted in proportion to $k/(\sqrt{w_i^{\rm data}}\,|r_i|)$, so that its contribution to the normal equations grows only linearly with the residual. We choose $k=3$, which is often called Huber weighting in the literature \citep{Huber_1964_AnnMathStat}. The weighting optimizes the Huber loss, which transitions from quadratic to linear at $3\sigma$. The Huber loss is convex, avoids modifying the uncertainties within $3\sigma$, and is well-suited for situations such as this one where we expect few outliers as opposed to a heavy-tailed outlier distribution. We repeat this weight update until the largest relative change in the fit linear coefficients is $<10^{-7}$ or 50 iterations are reached. A pixel is only fit if it has more measurements than basis vectors. Otherwise it is left empty and masked (Section~\ref{sec:mask}). For every pixel we store the fit amplitudes, their formal covariance from the inverse of the final (Huber-weighted) normal matrix, the reduced $\chi^2$, and a robust estimate of the scatter of the normalized residuals, which the masking stage uses. The line surface brightnesses we report are the integrals over frequency of the fitted feature basis vectors, in erg s$^{-1}$ cm$^{-2}$ sr$^{-1}$. The formal uncertainties are calibrated against the data in Section~\ref{sec:sens}.

\subsection{Other Contamination Removal} \label{sec:contam}

\begin{figure*}[t]
    \centering
    \includegraphics[width=\linewidth]{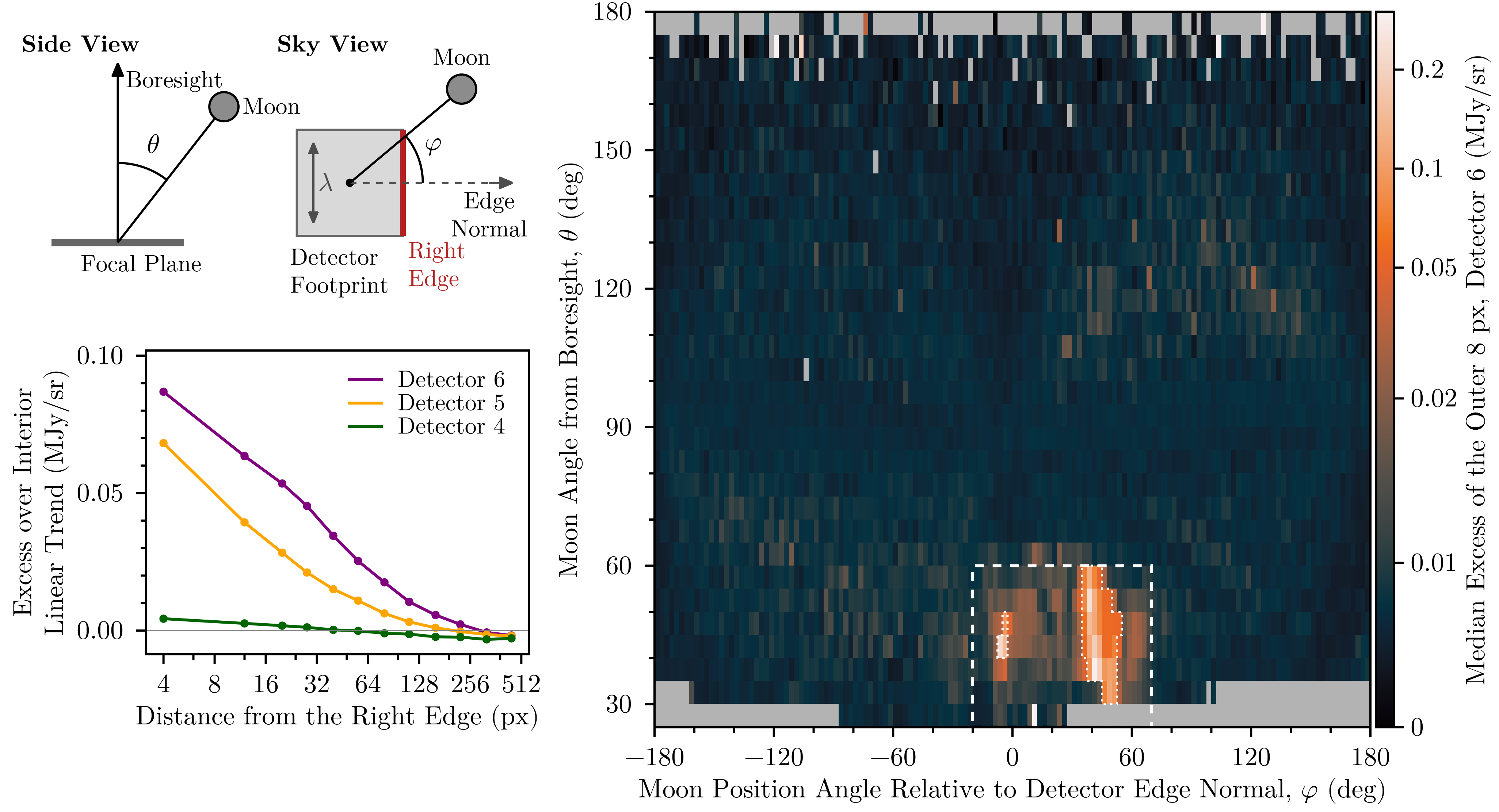}
    \caption{Identifying the effect of scattered moonlight on the right edge of the long-wavelength detectors. Left, top: Definition of the angle $\theta$ of the Moon from the boresight and the Moon's position angle $\varphi$ on the sky relative to the outward normal of the right detector edge. Left, bottom: Excess of a block of pixels at a given distance from the right edge over a linear trend fit to the interior pixels of the same wavelength (more than 128 pixels from either edge), for detectors 4, 5, and 6, as the median over the exposures in the core of the Moon lobe (dotted outlines: bins with a median excess above 0.05 MJy sr$^{-1}$ and at least 20 exposures) minus the median over 4,000 random exposures outside the geometric flag. Each point sits at the center of its block, and the blocks widen with distance from the edge, from 8 to 128 pixels wide. Right: The excess for the outermost 8 pixels of detector 6, for all 248,000 detector 6 exposures, binned by $\varphi$ ($2\fdg5$ bins) and $\theta$ ($5\degr$ bins) (gray indicates bins with $<5$ exposures). We use a logarithmic color stretch above 0.02 MJy sr$^{-1}$. The dashed rectangle is the geometric flag used in this work ($\theta<60\degr$, $\varphi$ from $-20\degr$ to $+70\degr$).}
    \label{fig:moon_edge}
\end{figure*}

Some forms of contamination are not easily detected in single-pixel fits. To handle them we defined two custom rejection methods described below for the most problematic cases.

\paragraph{Scattered moonlight} We find that one edge of each of the long-wavelength detectors (4, 5, and 6; 2.4--5.0\um) needs to be masked for some exposures. After investigating, we find that this edge-outlier occurrence depends strongly on the angular separation between the Moon and the boresight, and on the position angle of the Moon on the sky relative to the detector edge in question. The SPHEREx team has also noted this feature and attributed it to the Moon shining on the exterior of the instrument \citep{Bock_2026_ApJ, SPHEREx_ExpSupp_2026}. We compute the Moon direction for each exposure from the JPL DE440 ephemeris \citep{Park_2021_AJ} through SPICE \citep{Acton_1996_PSS}, at the exposure mid-time, from the spacecraft position recorded in the image headers, and the orientation of the detector edge on the sky from the image world coordinate system (WCS). In detector coordinates, this edge runs parallel to the direction of wavelength variation, at the ``right'' edge of the detector (pixels near the highest x-axis values), so the affected strip spans the full wavelength range of the detector. 

In Figure~\ref{fig:moon_edge}, we show the strength and angular dependence of this effect for every exposure of detector 6 in our dataset, measured as the excess of the outermost 8 pixels along this edge over a linear trend fit to the interior of the detector at the same wavelength. The excess appears only when the Moon is 30--60\degr from the boresight and lies on the side of this edge, in two narrow lobes of position angle, and it recurs every lunar month. We show the ramp profile of this excess as a function of distance from the edge for the exposures in the core of the lobe (Figure~\ref{fig:moon_edge}, bottom left). This shows that the effect is larger, and penetrates further in from the edge, for the longer wavelength detectors. This effect is about 20 times weaker on detector 4 than on detector 6 (0.004 versus 0.087 MJy sr$^{-1}$ in the outermost 8 pixels). For the exposures in the core of the lobe, the excess on detector 6 is still 0.01 MJy sr$^{-1}$ at 128 pixels and beyond about 300 pixels becomes indistinguishable from the interior linear trend fit. 

Until we better understand how to model this contamination, we mask the outermost 128 pixels from this edge on detectors 4, 5, and 6 (setting those measurements to have zero weight) for all exposures for which the angular separation to the Moon is less than 60\degr and the position angle of the Moon on the plane of the sky is within $-20\degr$ to $+70\degr$ of the outward normal of this edge (Figure~\ref{fig:moon_edge}, thick dashed box, right). This purely geometric flag encloses the lobe of the excess in Figure~\ref{fig:moon_edge} at the cost of 5.2\% of detector 6 exposures (4.4\% of the detector 4 exposures in our dataset). In the fields of this work it removes between 0.0\% and 1.3\% of the measurements, M31 (29 flagged exposures) being the most affected. Given the orientation, our hypothesis is that this is scattered moonlight that manages to leak in through the $\sim100$\um gap between the filter assembly and the detector \citep{Korngut_2018_SPIE} for these specific Moon orientations. 

\paragraph{Fast moving objects}  Single-pixel outlier flagging often struggles to identify faint moving objects because their per-pixel signal falls below detection thresholds even when the integrated trail is significant \citep{Whidden_2019_AJ, Nir_2018_AJ}. They then create coherent biases on the plane of the sky that are especially nefarious for studies of filamentary structure or image anisotropy. To mask pixels impacted by these faint moving objects, we do a first pass of the IRLS, and then run a Radon transform \citep{Deans_1983} on the per-exposure residual maps to identify linear streaks of outlying pixels. For each exposure and each \healpix $N_{\rm side} = 2^{7}$ ($\sim0\fdg5$) pixel of the fit, the normalized residuals $r_i\sqrt{w_i^{\rm data}}$ of that exposure are placed at their detector coordinates, median-subtracted, clipped at $\pm5$, and summed along lines of width 1--3 pixels and length 100, 200, 400 pixels, or the full footprint, at angles spaced by $0\fdg25$. Each sum, divided by the square root of the number of pixels it contains, is normalized by the median absolute deviation over all lines of the same width and length in that exposure, which gives a Radon peak score $S$ in robust standard deviations. 

Because the sky tiles cut long tracks into fragments, candidate fragments of the same detector line in the same exposure (within $1\degr$ in angle and 5 pixels in offset) are merged, with their scores summed and divided by the square root of their number. We impose a threshold of $S > 30$ on the merged Radon peak, and additionally require a total extent of at least 200 pixels along the line, a distance of at least 128 pixels from the detector edge (which separates tracks from the edge ramps above), and that the same sky pixels in the other exposures of the tile have a mean residual below 30\% of the residual on the track (which separates tracks from static features such as stellar wings or nebular edges). These thresholds were set using the two tracks found by eye. We then mask $\pm 3$ pixels around the line over the full extent of the detector, since the detected extent is only a lower bound on the length of the track (which may continue through tiles with too few measurements to be tested) and the strip costs only the equivalent of six columns of a single exposure. We then refit any \healpix pixels that used any of the measurements with weights now set to zero by the linear feature mask. Across the Rosette, Corona Australis, Cygnus X, Heart and Soul, M31, and M33 maps, we found 53 linear tracks in 49 of the 1,784 exposures and masked an additional 0.02\% of the measurements for this reason (0.001--0.06\% per field). In the $1\degr\times1\degr$ depth-test region at the north ecliptic pole (NEP), we found 237 tracks in 230 of the 6,183 exposures.

\subsection{Spectral Response of the LVF} \label{sec:LVF}

To forward model theoretical spectra to the SPHEREx measurements, we need to know the spectral response. Because SPHEREx uses a linear variable filter, this is not just a matter of looking up a static filter curve.\footnote{Though, in classical photometry it has been recognized for a long time that the apparent spectral response can vary across the field of view due to angle-dependence in the transmission function of the filter \citep{Lissberger_1959_JOSA, Doi_2010_AJ, Tonry_2012_ApJ, Li_2016_AJ}.} We instead need a spectral response model that varies across the detector. The SPHEREx team divides the instrument's wavelength coverage into 102 channels (17 per detector), and the ground spectral calibration of \citet{Hui_2026_ApJS} further divides each detector into 68 ``sub-channels'' of pixels that are co-spatial with respect to the variation in center wavelength of the linear variable filter, each with a measured spectral response. 

Instead of making this (sub-)channel approximation, we model the linear variable filter as continuous, reflecting its physical fabrication. We use the highest spatial resolution measurement of these sub-channels, the 68 measured sub-channel response profiles per detector of the ground spectral calibration \citep{Hui_2026_ApJS}, for the shape of the response only, and continuously interpolate the mapping of pixels to sub-channels, introducing decimal sub-channels. This allows us to define the spectral response of a given pixel on the detector by interpolating the measured (non-Gaussian) sub-channel spectral responses. We blend the two bracketing sub-channel profiles linearly, and the blend is shifted in wavelength so its half-transmission wavelength equals the center wavelength (band center) of that pixel in the public SPHEREx spectral WCS (the per-pixel \texttt{spectral\_wcs} calibration product of QR3 \citep{SPHEREx_ExpSupp_2026}, in the pixel convention of the Level 2 images). 

We test the efficacy of the interpolation by leave-one-out tests, predicting each measured sub-channel profile from its two neighbors, which is twice the separation of a real interpolation, so it is an upper bound on the interpolation error. We find average errors of 1.8\% of the peak transmission in the interpolated spectral response functions of detector 4, though only 0.5\% in the response integrated over the bandpass for a flat spectrum. The maximal errors are 12\% of the peak and 4\% in the bandpass-integrated response, which occur for the spectral response at the short wavelength edge of detector 4 (2.45\um), where the dichroic spectral response causes rapid variations in the effective spectral response. We distribute a package in Section~\ref{sec:dataavil} that implements this continuous spectral response model and can return the spectral response function of an arbitrary pixel in the focal plane in $\sim1$~$\mu$s having only loaded into RAM 49~MB of precomputed coefficients per detector (287~MB for all six).\footnote{This timing depends on the hardware and load. It is reported for a single thread on an otherwise idle Intel Xeon Platinum 8168 (2.7~GHz) node of the Flatiron Institute Popeye cluster, averaged over $2\times10^6$ random detector 4 pixels. A pixel whose maps are already in cache returns in 0.7~$\mu$s.}

When forward modeling narrow, well-characterized lines, we use this spectral response model to predict how these components will contribute to SPHEREx's measurements, for example for hydrogen recombination lines or molecular hydrogen ro-vibrational lines. In this case, the lines are well-known and the biggest source of modeling uncertainty is in the spectral response model. For the PAH emission, the shapes of the underlying spectral lines are not fundamentally well-determined and can vary. As a result, we learned data-driven corrections to basis vectors for our 3.3 and 3.4\um PAH components. In this case, we are not sensitive to the LVF in terms of the quality of our modeling of the data. The LVF only impacts our ability to do the under-determined task of deconvolving the data-driven basis vectors to infer a higher-resolution pre-filter-curve-convolution PAH spectrum, which we defer to future work.

Because the spectral response function model and wavelength solution are intimately coupled, our modifications to the spectral response model above could introduce wavelength offsets compared to the SPHEREx wavelength solution when forward modeling theoretical spectra. We do find some small wavelength offsets are necessary for detectors 3 and 4. The hydrogen recombination lines in the Rosette nebula are fit best when the response of each pixel sits 4~nm redward of its center wavelength on detector 4 and 2~nm redward in detector 3. This is about $4\%$ of the bandwidth in both bands. We apply these offsets to the narrow-line basis vectors by shifting their line wavelengths by the same amount in the opposite direction. We do not shift the PAH basis vectors, since we see that their data-driven corrections absorb the same offset. We have not attempted to determine whether these offsets reflect issues with the SPHEREx wavelength solution, the response profiles, or our interpolation of them.

\subsection{Models} \label{sec:models}

Our main spectral model focuses on making PAH maps. For this, we consider all observations within $\pm10$ bandwidths around both the 3.3 and 3.4\um PAH peak centers, all of which fall within detector 4. To model the continuum, we use three basis vectors: a constant, $\lambda^{2}$, and $\lambda^{-2}$, each normalized to unity at 3.3\um. These are motivated by our expectation of being in the Rayleigh-Jeans tail of starlight, which falls like $\lambda^{-2}$, and the presence of continuum emission from very small grains with a color temperature near 1000~K \citep{Sellgren_1984_ApJ}, which $\lambda^{2}$ approximates over detector 4. The three terms together describe the smooth continuum across our 2.5 to 3.9\um window with curvature of either sign. To handle regions where ionized gas is bright, we add one basis vector for hydrogen recombination emission. It contains all Brackett, Pfund, and Humphreys lines between 2.3 and 4.5\um together with the nebular continuum, with relative strengths fixed to Case B at $10^4$~K and $n_e = 10^2$~cm$^{-3}$ from PyNeb \citep{Luridiana_2015_AA, Storey_1995_MNRAS}. As a result, the hydrogen recombination basis vector has a single amplitude, which we report in terms of the corresponding Pf$\delta$ line surface brightness. We represent each line with a Gaussian of width 1.5~nm, far narrower than the spectral response, and we shift this basis vector by $-4$~nm to compensate for a detector 4 wavelength offset we find necessary to optimize the goodness of fit for these narrow lines (see Section~\ref{sec:LVF}). 

\subsubsection{Data-Driven Corrected PAH Basis Vectors} \label{sec:ddPAH}
We found that simple analytic forms for the PAH emission were insufficient to describe the data well and left significant banding in our maps. To solve this problem, we initialized our PAH basis vectors using higher resolution spectra of PAH emission from JWST NIRSpec, obtained by the PDRs4All program toward the Orion Bar \citep{Berne_2022_PASP, Peeters_2024_AA}. We use the published template spectrum of the atomic PDR region from the PDRs4All website.\footnote{\url{https://pdrs4all.org}} We then separated the 3.3 and 3.4\um emission into two basis vectors. After masking the emission lines listed by \citet{Peeters_2024_AA} and subtracting a linear continuum, we fit their Gaussian decomposition to the spectrum, which includes Gaussians at 3.246, 3.29, 3.33, 3.40, 3.42, 3.46, 3.52, and 3.56\um plus a broad 3.42\um plateau. We use these Gaussians only to apportion the emission. At each wavelength, the observed spectrum is divided between the two basis vectors in proportion to the Gaussians of the 3.3\um group (3.246, 3.29, 3.33\um) and the 3.4\um complex (3.40, 3.42, 3.46, 3.52, 3.56\um), so the basis vectors keep the observed line shapes rather than a combination of Gaussians approximation. We keep no separate plateau component and its emission is shared between the two templates in the same proportion. 

Each template is normalized by its integral in frequency, so the fitted amplitude is the integrated surface brightness of the whole feature. These two spectra are our initial set of PAH basis vectors. Because of some combination of issues with the spectral response, wavelength solution, or difference in the average shape of PAH emission compared to that PDR, we still observe some banding in our PAH maps when using these basis vectors. As a result, we learn data-driven corrections to these basis vectors on five training fields: $2\degr$ boxes toward $\rho$ Ophiuchi, the Rosette nebula, Chamaeleon I, NGC 7023, and IC 348, chosen to span ecliptic latitude, scan angle, and surface brightness. Five more fields (Lupus I, NGC 2068, TMC-1, the Polaris flare, and the Ursa Major cirrus) were held out, and we used the amplitude of banding in those maps to set hyperparameters and refine the procedure presented here. We make the resulting basis vectors public in Section~\ref{sec:dataavil}.

We learn the corrections in the space of the measurements rather than on a wavelength grid. For PAH basis vector $m$ (the 3.3 or the 3.4\um basis vector) and a measurement whose spectral response is centered at wavelength $\lambda$, the model value is
\begin{equation} \label{eq:learn_ds}
    B_m(\lambda) = \mathcal{F}\!\left[T_m^{\rm init}\right](\lambda) + \sum_j C_{jm}\, \beta_j(\lambda),
\end{equation}
where $\mathcal{F}[T_m^{\rm init}]$ is the initial basis vector convolved with the response of that measurement, $\beta_j$ is the $j$th cubic B-spline in $\lambda$ on a grid of knots 10~nm apart, and $C_{jm}$ is the coefficient of spline $j$ for basis vector $m$. The sum runs over the splines that lie entirely within the range in which we allow corrections, 3.04--3.50\um for the 3.3\um basis vector and 3.24--3.72\um for the 3.4\um basis vector, 43 and 45 splines respectively, each spanning 40~nm. The model for measurement $i$ of map pixel $p$ is then $\sum_m a_{pm} B_m(\lambda_i)$ plus the continuum, where $\lambda_i$ is the center wavelength of that measurement's response. The correction is thus a smooth function of wavelength with 88 free coefficients in total. The coefficients $C$ are the same for every map pixel and every field with only the amplitudes $a_{pm}$ varying from pixel to pixel. 

We find $C$ by minimizing the Huber loss, with two added regularization terms,
\begin{equation} \label{eq:learn}
    \sum_{p}\sum_{i \in p} w_i^{\rm new}\, r_i^2 + s_1\,\|C\|^2 + s_2\,\|D^2 C\|^2
\end{equation}
while optimizing both the per-pixel amplitudes and $C$. Here $p$ runs over the training map pixels and $i$ runs over the measurements of pixel $p$. The residual $r_i$ of measurement $i$ is relative to the model of Section~\ref{sec:linfit} with its PAH columns defined by Equation~\ref{eq:learn_ds}, and $w_i^{\rm new}$ is its weight after the IRLS of Equation~\ref{eq:irwls} has converged, so the first term is the same Huber-weighted loss that the production map-making code minimizes. To this, we also add a ridge term that encourages smaller corrections and a penalty that encourages smoothness along the knot sequence,
\begin{equation} \label{eq:learn_pen}
\begin{aligned}
    \|C\|^2 &= \sum_m \sum_j C_{jm}^2, \\
    \|D^2 C\|^2 &= \sum_m \sum_j \left(C_{j+1,m} - 2C_{jm} + C_{j-1,m}\right)^2,
\end{aligned}
\end{equation}
where the second is the squared second difference of the spline coefficients. The hyperparameters chosen for the amplitude of these terms are $s_1 = 1\times10^{-3}$ and $s_2 = 1\times10^{-1}$, respectively. Each correction is constrained to integrate to zero, $\sum_j C_{jm} \int \beta_j\, d\nu = 0$ for each basis vector $m$, which we impose via a Lagrange multiplier in solving for $C$, so the normalization of the initial basis vectors is preserved. The penalty strengths and the wavelength ranges of the corrections were chosen by scoring the maps of the five held-out fields with a model-independent banding metric, discussed below, breaking ties using the best $\chi^2$.

We alternate between the per-pixel IRLS fits with $C$ fixed and a weighted least-squares update of $C$ with the amplitudes and the weights $w_i^{\rm new}$ fixed, on every third unmasked pixel of the training fields (2.9 million pixels), until the largest change in $C$ is below $10^{-3}$. The robust $\chi^2$ per degree of freedom of the training pixels falls from 1.40 with the initial basis vectors to 1.35 with the corrections. 

Because the corrections of Equation~\ref{eq:learn_ds} are defined only through the LVF, we do a final solve for smooth, non-negative basis vectors on the 2.5~nm wavelength grid shared by the rest of our basis vectors that reproduce the $B_m$. This is a quadratic program that matches each $B_m$ through the response of every training measurement, keeps the integral of the initial basis vectors, and penalizes curvature and distance from them. We take the smoothest solution within $5\times10^{-4}$ of the best $\chi^2$ per degree of freedom. While we infer smooth, high-resolution PAH basis vectors, we caution against interpreting their high resolution structure. We only show and use the LVF-convolved PAH basis vectors. Understanding how to deconvolve the data-driven corrections and structure we find goes beyond the scope of this work. Conceptually, deconvolution always requires a regularization choice, though one possible route would be to regularize based on a sparse sampling of higher resolution spectra.

Because Equation~\ref{eq:learn} is the data likelihood of the training fields, our data-driven corrections improve the $\chi^2$ of our fit by construction. As an independent check, we use a banding metric that we never optimize directly. It asks how much of the small-scale structure in a map can be predicted from the wavelength sampling alone. To measure it, we compute for every map pixel seven descriptors of how its measurements sample the spectrum, which serve as regressors: (1) the fraction of the measurements whose center wavelength lies within about one bandwidth of each of three reference wavelengths (3.29, 3.40, and 3.46\um), (2) the mean offset of those measurements' center wavelengths from the 3.29 and 3.40\um peaks, in units of the bandwidth, and (3) the signed offset of the nearest measurement from the same two peaks. We bin our PAH maps and these descriptors on a $14\farcs4$ grid, remove structure larger than $6\arcmin$ from both using a high-pass filter, and divide the map by its smooth version, so that the result is a fractional fluctuation. We then try to predict the map of interest from the regressors. We take the root-mean-square of the model fit using these regressors as our measure of the banding strength, the fraction of the local surface brightness that the sampling pattern alone explains. Because real structure can correlate with the descriptors by chance, we repeat the regression with the descriptor maps reflected, rotated, and shifted, and subtract the median of these null results in quadrature, so a map with no banding scores zero within the noise of the metric. We tracked these scores on both the summed $3.3+3.4$\um surface brightness and the $3.4/3.3$\um ratio maps.

For the Rosette nebula, adding the data-driven corrections to our JWST PDRs4All initialization halves the banding in the $3.4/3.3$\um PAH map, from 4.7\% to 2.2\% of the local surface brightness. The banding of the summed $3.3+3.4$\um map in Rosette is $\sim2\%$ and only marginally detected with either set of basis vectors. Relative to the initial basis vectors, those with data-driven corrections reduce the banding in every field and in both maps. Across the five training and three bright held-out fields, the banding of the summed $3.3+3.4$\um surface brightness drops from 0.9\% to 0.6\% of the local surface brightness. At that level it is only marginally detectable against the null tests ($1.5\sigma$). The banding of the $3.4/3.3$\um ratio in the same fields halves, from 1.8\% to 0.9\%. It remains detectable ($8\sigma$), driven by the Rosette and TMC-1 fields, and the three bright held-out fields alone show the same halving but remain $5\sigma$ detectable. On our two faintest cirrus test fields, the Polaris flare and the Ursa Major cirrus, the data-driven corrections reduce the banding in the ratio maps from 3.0\% to 2.0\%. This is still well above the null tests ($6\sigma$). These fields are the hardest case for a ratio map. Both features are faint, so the per-pixel ratio divides noise by noise, and sampling-dependent biases that are negligible against bright emission become a measurable fraction of the signal. Improving our model sufficiently to bring banding below the detection threshold in all fields will be the subject of future work, and may require higher dimensional representations of the PAH features and learning the basis vectors jointly with the map (at inference time) to capture spectral shape variation that correlates with environmental changes. For now, visual comparisons of our maps with others in the literature (see Section~\ref{sec:compare2others}) demonstrate the substantial improvement we have achieved in reducing banding.

\begin{figure}[t]
    \centering
    \includegraphics[width=\linewidth]{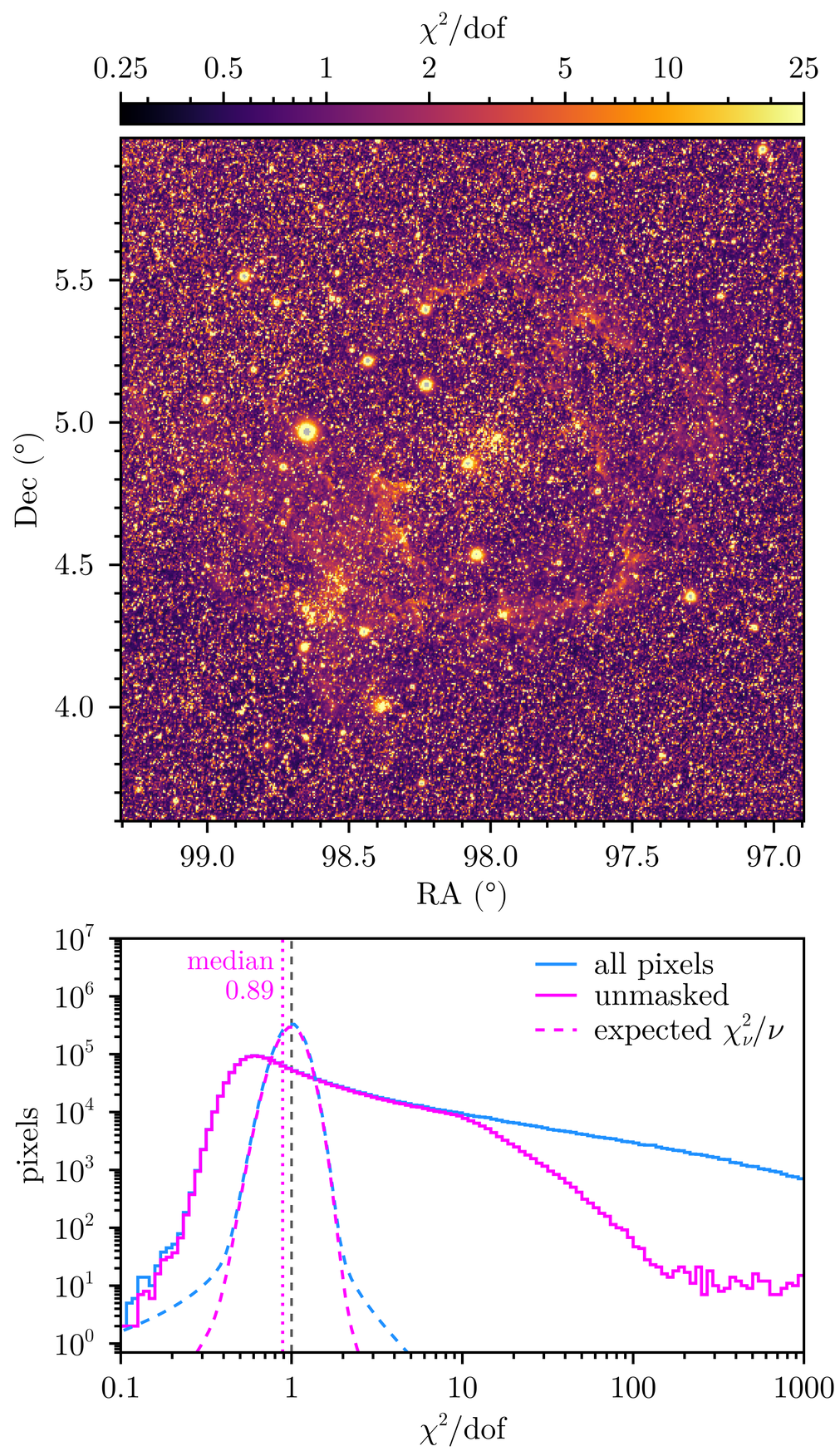}
    \caption{Goodness of fit of our linear forward PAH model for the Rosette nebula. Top: Map of $\chi^2/{\rm dof}$ of each pixel, using the pipeline uncertainties without a model-error floor. Here the dof is the number of measurements contributing to a pixel minus the six model parameters, which has a median of 75, and $\chi^2$ is simply summed over all of the measurements contributing to the pixel. Pixels masked for any reason (Section~\ref{sec:mask}) are gray. Bottom: Histogram of $\chi^2/{\rm dof}$ for all pixels (blue) and for unmasked pixels (magenta). The 
    dashed curves show the histograms expected if the $\chi^2$ of every pixel followed a $\chi^2$ distribution given its own number of dof. The dotted line marks the unmasked median of $0.89$ and the dashed line marks $\chi^2/{\rm dof} = 1$.}
    \label{fig:chi2dist}
\end{figure}

\subsubsection{Fit Quality} \label{sec:fitqual}

In Figure~\ref{fig:pixel_schematic}, we show the fit of this model to an example pixel in the Rosette nebula. In the left panel, we demonstrate which measurements are included, the spatial weighting scheme, and how the central wavelength of the measurements varies spatially. In the right panel, we show how the basis vectors combine to give the total model and compare the model to the data, with the uncertainty normalized residuals shown in the bottom subpanel. Here, the uncertainties we show include the final multiplicative updates to the weights from the IRLS procedure. We provide markers showing the locations of the center wavelength of the main PAH features, as well as the three strongest hydrogen recombination lines in the wavelength range (Br$\beta$, Pf$\gamma$, and Pf$\delta$). The Pf$\delta$ feature directly overlaps the PAH features.

The right panel of Figure~\ref{fig:pixel_schematic} also shows the average wavelength spacing at the current point in the mission. Because the phase of the wavelength sampling is fairly random away from the ecliptic poles, our ability to learn and constrain spectral substructure should improve with increasing duration of the SPHEREx mission, because it will bring an increase in the spectral sampling density. From this perspective, the deep but uniform wavelength sampling of the ecliptic poles is a missed opportunity for data-driven refinements of the spectral response model of SPHEREx and pushing the limits of the spectral resolution of the components we can infer from the data.

To demonstrate how consistent our model is with the data on a larger scale, we show the $\chi^2$ distribution for the Rosette nebula testbed in Figure~\ref{fig:chi2dist}. After excluding masked pixels from our fit (e.g. stars, see Section~\ref{sec:mask}), we find a median $\chi^2/{\rm dof} = 0.89$. This suggests that our model is describing the data well on average. The fact that our typical $\chi^2/{\rm dof}$ is less than one might reflect that we added in too much freedom to our model, for example in terms of the number of continuum components. However, looking at the overall distribution of the $\chi^2/{\rm dof}$, we see that it is extremely heavy-tailed compared to the statistically expected distributions. This heavy-tailing is especially prominent in the direction of pixels with elevated $\chi^2/{\rm dof}$. We take this bottom panel as a strong indication that we are not fully describing the complexities of the SPHEREx observations, suggesting there is an opportunity to extract much more information from the data. While it can be disheartening to see that your model is misspecified, having access to the residuals is one of the key benefits to a forward model, and is not accessible to even advanced channel mapping methods \citep{Hora_2026_ApJ}.

In the top panel of Figure~\ref{fig:chi2dist}, we show the plane-of-sky map of $\chi^2/{\rm dof}$. The relative paucity of spatial structure in this map acts as further validation of our model. In terms of spatially correlated structure, we do see $\chi^2/{\rm dof} \sim3$ along the ridges of brightest PAH emission. This could indicate that the PAH lineshape varies in regions of extremely high UV flux, which we speculate may be because these higher UV fluxes stimulate vibrational hot bands \citep{Barker_1987_ApJL} or modify the underlying carrier distributions.

In addition to the model described above, we also consider two other models used in this paper. For some of the test fields where we do not expect significant emission in hydrogen recombination lines, we simply use the model described above, but without the hydrogen recombination basis vector. We also consider a pure hydrogen recombination line model that we apply only to the strongest lines in the SPHEREx wavelength range. We query $\pm4$ bandwidths around Pa$\alpha$ (1.876\um), which is a subset of detector 3 that also contains the Br$\delta$ (1.945\um) and Br$\epsilon$ (1.818\um) lines. We model the three lines as a single basis vector with Case B line ratios \citep{Hummer_1987_MNRAS, Luridiana_2015_AA} and report the fit amplitude in terms of the Pa$\alpha$ line surface brightness. As above, we represent each line with a Gaussian of width 1.5~nm, shifted here by $-2$~nm to account for our apparent detector 3 wavelength offset (see Section~\ref{sec:LVF}). For the continuum, we use the same three basis vectors as above, a constant, $\lambda^{2}$, and $\lambda^{-2}$ term, here normalized to unity at 1.876\um.

\subsection{Masking} \label{sec:mask}

Not every pixel fit is a measurement of diffuse emission. After fitting every \healpix $N_{\rm side} = 2^{15}$ pixel, we mask those pixels where the fit likely captures something else, such as a star or a poorly modeled spectrum. The pixel mask is applied after the fit and is distinct from the measurement-level rejections of Section~\ref{sec:contam} (Moon-contaminated detector edges and moving-object tracks, which enter the fit with zero weight) or IRLS down-weighting of outliers. The mask involves six different cuts applied in the order listed. The cuts were designed around simple principles and evolved over time in response to by-eye inspection of our test fields. These cuts are admittedly somewhat contrived and this is one of our primary motivations for modeling both the diffuse emission and point sources at the same time, as discussed in Section~\ref{sec:conc}.

\paragraph{Coverage} The fit itself requires more measurements than basis coefficients for the linear system to not be singular. For models including PAH components, we also mask pixels with three or fewer measurements with a central wavelength within one bandwidth of $3.29$\um. This ensures we have sufficient sampling of our spectral feature of interest. For the Pa$\alpha$ fit, we apply the same rule about $1.876$\um. This cut is mild and removes less than $0.6\%$ of every field.

\paragraph{Star} We seed the star list with unWISE sources brighter than $W1 = 9.5$ mag \citep{Schlafly_2019_ApJS}. Where unWISE is saturated ($W1 < 8$ or flagged), we use the 2MASS $K_s$ magnitude of the counterpart within $3\arcsec$ instead \citep{Skrutskie_2006_AJ}, and we add 2MASS stars brighter than $K_s = 8$ mag that unWISE lacks. To predict how far each star contaminates the line map, we measure its continuum halo and the fraction of that halo that leaks into the line map. We model the halo with a radial template $h(r) \propto [1 + (r/r_c)^2]^{-\beta/2}$, which we fit to the stacked halos of the field's bright stars. For fields with fewer than 100 bright stars with measurable halos, we use the median template of the other fields instead. For each star we then measure the amplitude $A$ of this template on the fit $3.29$\um continuum map. The leakage fraction $F$ comes from a regression of the stacked line-map profiles on the stacked continuum profiles, taking the larger of the $3.3$ and $3.4$\um values since we use one mask for both maps. 

In nebular regions, part of the continuum excess around a star is ISM emission rather than halo. Where the $I_{3.3}$/continuum color of the halo region is closer to the ISM than to a star, we split the excess into a stellar part with color $F$ and an ISM part with the local ISM color, and use only the stellar part when fitting $A$. The star is masked out to the radius where its predicted contamination of the line map, $A\,F\,h(r)$, falls below $0.2\max(\sigma_I, 0.2\,I_{\rm loc})$, with $\sigma_I$ and $I_{\rm loc}$ the scatter and median of $I_{3.3}$ in an annulus around the star, respectively. The annulus we use scales with the brightness of the star, spanning $3$--$4\,R_s$, where $R_s = 10^{2.33 - 0.082\,m}$ arcsec and $m$ is the seed magnitude, with a floor of $120\arcsec$--$200\arcsec$ that applies to all stars fainter than 9 mag.

For stars that leave a coverage hole of radius $r_{\rm hole}$, for example due to saturation, or are brighter than $6.5$ mag, we then grow the mask outward in $15\arcsec$ rings. Here $r_{\rm hole}$ is the radius of the star's coverage hole or, where larger, of a connected patch of at least 30 cut pixels around it. At each step we compare the median $I_{3.3}$ of the $15\arcsec$ ring just outside the current disk with that of the $60\arcsec$-wide reference annulus beyond it, using only unmasked pixels attributed to this star. The growth continues while the ring deviates from the reference annulus by more than twice the standard error of the difference, estimated from the robust scatter in the reference annulus, and by more than half that scatter. The second condition prevents small but statistically significant gradients from growing the disks of stars in bright regions. We grow on deviations of either sign, since the leaked continuum can be oversubtracted or undersubtracted. If a ring is brighter than the reference annulus and the local ISM color can be measured, we also confirm that the ring has a continuum excess and that the ratio of the $I_{3.3}$ excess to the continuum excess is below half the local ISM ratio. This check stops the growth when the emission is dominated by real diffuse emission, which has the local ISM color. We cap the radius of each star mask at the larger of $60\arcsec$ and $3 r_{\rm hole}$. For stars brighter than $6^{\rm th}$ mag, we make the cap magnitude dependent and it instead grows with brightness as $60\arcsec\,10^{0.2(6 - m)}$, which doubles every $1.5$ mag ($380\arcsec$ at $2^{\rm nd}$ mag). For the few stars brighter than $2^{\rm nd}$ mag, we relax the cap to $0\fdg5$.

\paragraph{Residuals} For each pixel the fit stores the robust scatter (interquartile range divided by 1.349) of the normalized residuals $r_i/\sqrt{\sigma_i^2 + (f_{\rm m}\, m_i)^2}$ over the measurements shortward of $3.36$\um, where $m_i$ is the model prediction and $f_{\rm m}$ a per-measurement model-error floor. We mask pixels where this scatter exceeds 4. This restriction to only use the shortward side is an attempt to reduce sensitivity to detailed modeling of the 3.4\um feature shape. The floor is calibrated per field on star-free pixels as the value at which the residual scatter becomes independent of the model signal-to-noise. It ranges from $0.05$ (Cygnus X) to $0.18$ (Heart and Soul). The uncertainty on each value is $0.003$--$0.018$, based on a spatial jackknife that compares different regions of the image. In two fields, Corona Australis and the NEP, the calibration does not bound $f_{\rm m}$ below $0.2$. For these we use the median of the other fields, $0.12$. 

Fainter stars not in the seed list are caught by a second residual cut. The floor $f_{\rm m}$ can hide a poorly fit point source. Near a star the model $m_i$ includes the star's continuum, so the floor term $f_{\rm m} m_i$ grows with the star's brightness. A misfit of order the star's signal can then leave the robust scatter below 4 with the floor while far exceeding 4 without it. As a result, we also compute the residual scatter without the floor and mask pixels where it exceeds 4, but only for pixels within $9\arcsec$ of an unWISE $W1 < 16$ source that is bright enough to matter, meaning its expected continuum peak from the field's $W1$-to-continuum scaling exceeds four times the floor term $f_{\rm m} |m|$. 

Sources whose core has a color consistent with diffuse emission are exempt from this second residual cut. To evaluate the color, we compare the median $I_{3.3}$ and continuum of the pixels within $6\farcs5$ of the source with those in a $12\arcsec$--$24\arcsec$ annulus. The source is exempt when its core continuum excess is significant, at three times the annulus continuum scatter, and the ratio of the $I_{3.3}$ excess to the continuum excess exceeds half the field's median $I_{3.3}$/continuum ratio. This check works because a star adds continuum but no $3.3$\um PAH emission, so its PAH/continuum ratio is near zero, in contrast to a compact \ion{H}{2} region or PAH knots. This exception protects, for example, the \ion{H}{2} regions of M33. 

Finally, because the $3.4$\um map is more sensitive to stellar continuum leakage than the $3.3$\um map, the disks of stars brighter than $W1 = 6$~mag are grown further using the $3.4$\um map. At each step we compare the $15\arcsec$ ring just outside the current disk with the $60\arcsec$-wide reference annulus beyond it, using only unmasked pixels. In both we measure $q = I_{3.4} - \rho I_{3.3}$, where $\rho$ is the ratio of the median $I_{3.4}$ to the median $I_{3.3}$ in the reference annulus. The idea is that $q$ vanishes for emission with the local ISM color, so it isolates stellar excesses. We define $\Delta q$ as the median $q$ of the ring minus the median $q$ of the reference annulus, and estimate its standard error from the robust scatter of $q$ in the reference annulus. The disk is grown by $15\arcsec$ while $|\Delta q|$ exceeds both twice its standard error and $5\%$ of the median $I_{3.4}$ of the reference annulus. The second condition is an attempt to stop the growth on bright backgrounds. There, even small color gradients are statistically significant and would otherwise keep the disk growing. We cap this growth at $\min(4 R_0, R_0 + 300\arcsec)$, where $R_0$ is the disk radius before this step. As a final step, we mask any unmasked pixel with at least three of its four neighbors masked, and repeat until no pixel changes. This closes the isolated holes that the residual cuts leave at the cores of compact sources.

The masked fraction of each displayed field is $2.3\%$ (M33), $5.4\%$ (M31), $6.9\%$ (NEP), $11.0\%$ (Rosette), $15.0\%$ (Heart and Soul), $15.3\%$ (Corona Australis) and $34.5\%$ (Cygnus X). In the Galactic plane fields, the two residual cuts dominate. For the Pa$\alpha$ map of the Rosette nebula, we apply the mask of the PAH fit in addition to its own coverage cut. In the released maps, the mask is a bit field, so a user can tell which cut masked each pixel.

\begin{figure*}[t]
    \centering
    \includegraphics[width=0.9\linewidth]{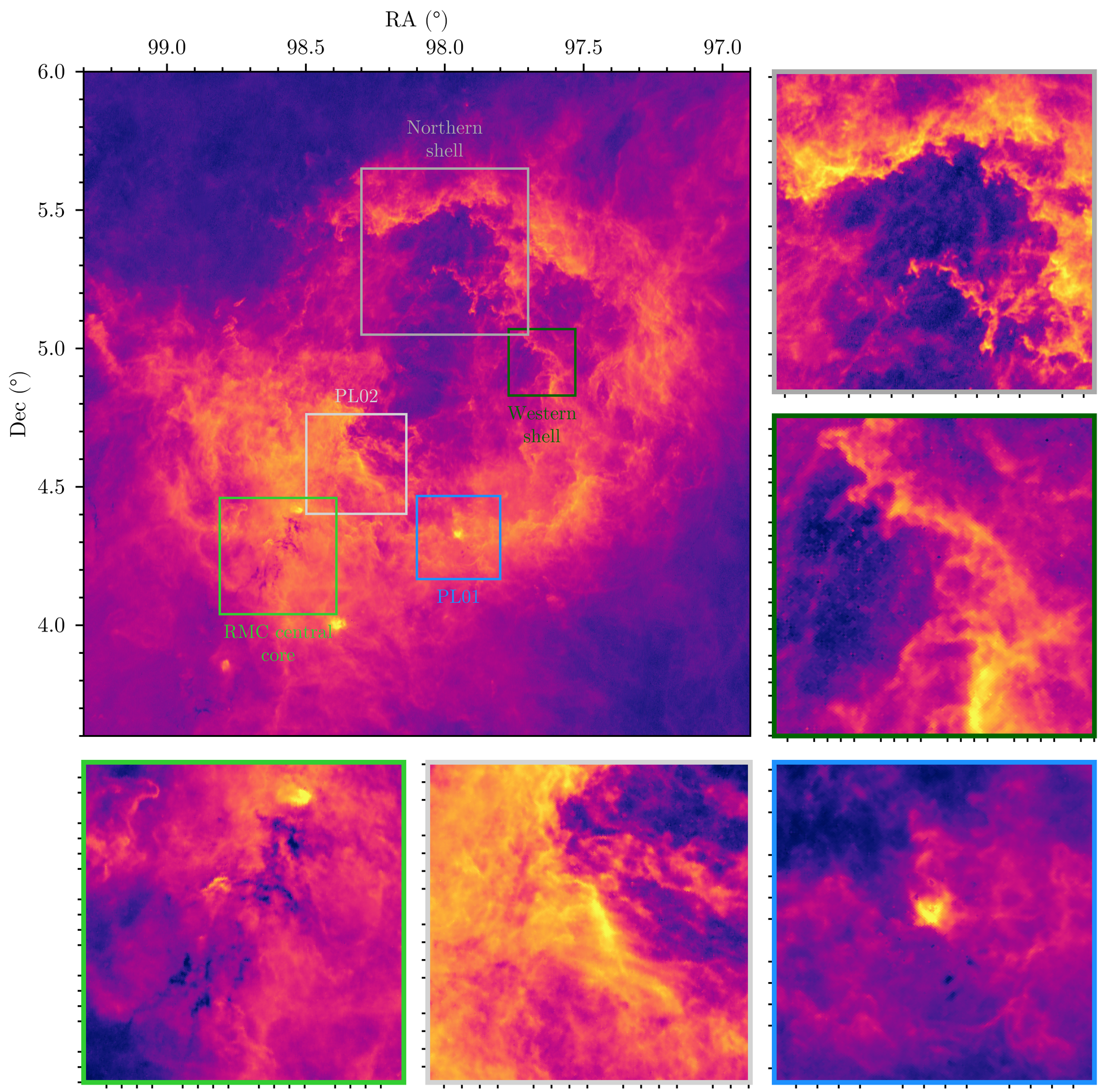}
    \caption{Overview of a few regions of interest in the Rosette nebula indicating their location on our full Rosette image and providing zoom-in subpanels for each region: the northern shell (top right), the western shell (middle right), and, along the bottom row from left to right, the central core of the Rosette molecular cloud \citep[RMC;][]{Blitz_1980_ApJ}, and the embedded clusters PL02 and PL01 \citep{Phelps_1997_ApJ,RomanZuniga_2008_ApJ}. Each subpanel has an independent log color scale from the 0.1--99.9$^{\rm{th}}$ percentile, with brighter colors representing more emission.}
    \label{fig:rosetteMulti}
\end{figure*}

\subsection{Infilling} \label{sec:infill}

To create clean maps with statistically consistent values in the masked pixels described above, we infill them with a data-driven, anisotropic, conditional Gaussian process method that is a slightly modified version of the method presented in \citet{Saydjari_2022_ApJ}. Briefly, we use unmasked pixels as ``training data'' of what the structure in an $N_p \times N_p$ pixel patch looks like ($N_p = 31$ pixels, $3\farcm3$) over a surrounding region of $\rm{widx} \times \rm{widx}$ pixels ($\rm{widx} = 81$ pixels, $8\farcm7$). The $5{,}101$ shifted copies of the patch within that region serve as samples from which we build the $N_p^2 \times N_p^2$ pixel-pixel covariance matrix, which does not impose any isotropy or functional form, but instead simply measures the correlations of pixels for each set of 2-dimensional pixel separations. 

Then, for the $N_p \times N_p$ patch centered on a masked pixel, we predict the values of all masked pixels in the patch at once, conditioned on the unmasked pixels of the patch, using this covariance matrix as the prior. We store the conditional mean, which is what the maps in this work show, two independent draws from the conditional distribution, and a per-pixel standard deviation, which we report as the uncertainty on infilled pixels in place of the measurement error. Because pixels infilled by earlier patches are treated as known when conditioning later ones (the mean on their means and each draw on its own earlier draw, so that the draws remain a consistent sequential simulation), the conditional variance of a single patch underestimates the uncertainty, by up to $10\%$ in our most heavily masked field. We therefore add the variance propagated from the earlier-infilled pixels in the conditioning set, approximating their posterior covariance by their stored variances and the sample correlation of the patch. Checked against 32 conditional draws, the resulting standard deviation is calibrated to a few percent. Regardless, the uncertainty we report for infilled pixels is a prediction uncertainty rather than a noise estimate so the mask column of the released maps identifies which pixels are infilled and carry this kind of uncertainty. The choice of $N_p$ is driven by computational cost and the choice of $\rm{widx}$ is driven by the spatial scale at which the structure in the image is changing, the scale at which approximating the samples as stationary breaks down, and the need to have enough samples to stably estimate the $N_p^2 \times N_p^2$ covariance matrix (here $5.3$ samples per matrix dimension).

Because we are infilling one patch at a time, and pixels infilled earlier are treated as known when conditioning later patches, the order of choosing these patches introduces some determinism to the map, especially when the masking fraction is large or there are large contiguously masked regions. We infill the masked pixels within the displayed region of each field in independent tiles of $0\fdg75$, each with $0\fdg2$ padding and a fixed random seed, so that the result is reproducible and tiles can be run in parallel. Within a tile the masked pixels are visited in the order of the NESTED \healpix index, which is a hierarchical space-filling curve, so that the fill proceeds through compact sub-regions in turn rather than sweeping across the tile in a single direction.

The major change compared to \citet{Saydjari_2022_ApJ} is adapting this algorithm to work on a \healpix grid as opposed to a rectilinear grid. For RING-ordered \healpix maps in the equatorial zone, below absolute latitudes of $41\fdg8$, every ring holds the same number of pixels ($4 N_{\rm side}$) and consecutive rings are offset by half a pixel, so an $N_p \times N_p$ patch and its shifted copies are a fixed pattern of index offsets. We previously used this trick to infill masked pixels in the DECaPS2 dust map \citep{Saydjari_2025_ApJ}.

To handle the higher latitudes where the number of pixels per ring changes, we rotate each field so that its center lies on the equator of a new \healpix frame, resample the map onto that frame (each rotated pixel takes the value of the original pixel containing its center), infill there, and carry the infilled values back to the original \healpix pixels. Since our fields currently extend at most a few degrees from their centers, the whole field lies deep within the equatorial zone of the rotated frame regardless of its declination. Because the two grids are different tessellations wherever the field is off the equator, a rotated pixel is treated as masked both when its center lies in a masked pixel and when it is the pixel that will supply the value of a masked pixel on the way back, so that every masked pixel receives a prediction (with its standard deviation) rather than the value of a neighboring unmasked pixel. 

\begin{figure*}[t]
    \centering
    \includegraphics[width=\linewidth]{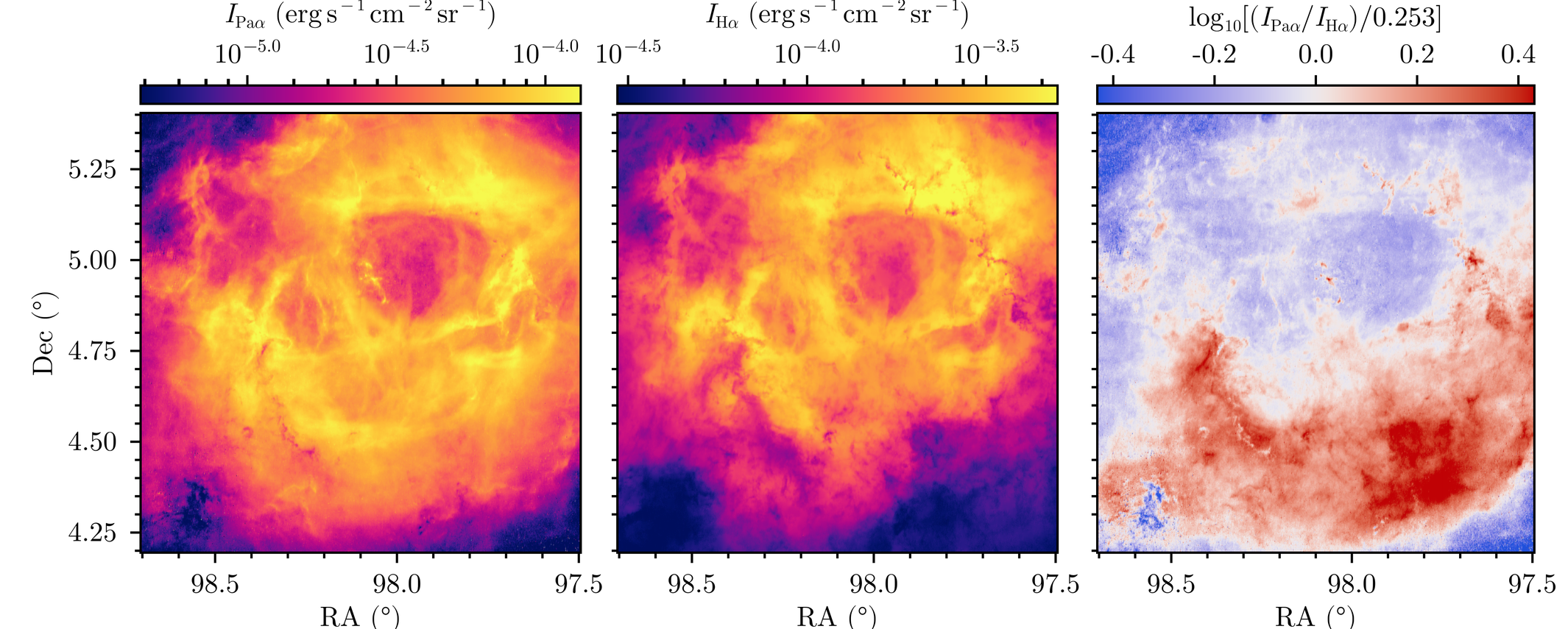}
    \caption{The LVFit Pa$\alpha$ SPHEREx map (left) compared with H$\alpha$ from the star-removed MDW H$\alpha$ Sky Survey DR1 mosaic \citep{Aftab_2025_arXiv} (center), and their ratio $\log_{10}[(I_{{\rm Pa}\alpha}/I_{{\rm H}\alpha})/0.254]$ (right), where 0.254 is the median ratio over the displayed region. The H$\alpha$ surface brightness uses the nominal MDW conversion of 86.7\,R per MJy\,sr$^{-1}$ and includes [\ion{N}{2}] and continuum contributions, so its absolute level is uncertain at the $\sim10$\% level. Red regions toward the Rosette molecular cloud and along dark globules indicate enhanced extinction of H$\alpha$. The MDW stacks have a PSF FWHM of $\sim6.4$--$8\farcs2$ (median over stars in the four fields used), comparable to the $8\farcs5$ FWHM angular resolution of our map.}
    \label{fig:rosette_pa_halpha}
\end{figure*}

\section{Galactic Testbeds} \label{sec:galactic}

\subsection{Rosette Nebula} \label{sec:rosette}

The Rosette nebula is a well-known \ion{H}{2} region that is ionized by OB stars in the star cluster NGC 2244, at a distance of $\sim1.4~{\rm kpc}$ \citep{Lim_2021_AJ}. The intricate structure of the nebular emission provides an ideal testbed for our methodology, and allows us to investigate the detailed morphology of the 3.3\um and 3.4\um emission across a region that is illuminated by well-characterized photoionization sources. 

Figure~\ref{fig:rosette_methodcompare_full} shows the LVFit 3.3\um PAH line surface brightness map of the Rosette nebula, along with the surface brightness of the LVFit continuum model at 3.29\um. For comparison, we show a simple band-averaged map of the same region at 3.3\um. This band-averaged, continuum-subtracted map is made by taking the inverse-variance weighted average of all SPHEREx observations in a given \healpix pixel within two times the LVF bandwidth of the PAH central wavelength of 3.29\um minus the average of observations between two and five bandwidths away. The naïvely band-averaged map showcases the severe banding artifacts discussed above. The LVFit 3.3\um PAH line map shows a wealth of detailed morphological structure, and we highlight our spatial resolution with a zoomed-in version of Figure~\ref{fig:rosette_methodcompare_full} in Figure~\ref{fig:rosette_methodcompare_small}. Adopting a distance to the Rosette nebula of 1.4 kpc, the physical resolution of our maps ($8\farcs5$ FWHM) is $\sim0.06~{\rm pc}$. We also highlight several regions of interest in Rosette discussed in the literature in Figure~\ref{fig:rosetteMulti}. 

\subsubsection{Paschen-$\alpha$} \label{sec:paa}

Understanding the multiphase spatial structure of the \ion{H}{2} region is essential for interpreting the observed morphology of PAH emission. SPHEREx observations can also trace the warm ionized medium at the center of the \ion{H}{2} region via hydrogen recombination lines, the strongest of which in the SPHEREx wavelength range is Pa$\alpha$. Notably, Pa$\alpha$ is not accessible from the ground due to water vapor absorption as it falls between the H and K infrared atmospheric windows. In Figure~\ref{fig:rosette_pa_halpha}, we show the LVFit SPHEREx Pa$\alpha$ line surface brightness map of the Rosette nebula made by jointly modeling Pa$\alpha$, Br$\delta$, and Br$\epsilon$. While Pa$\alpha$ is not accessible from the ground, we can compare our map to recent high angular resolution maps ($\sim6.4$--$8\farcs2$ FWHM) of H$\alpha$ from the MDW H$\alpha$ Sky Survey \citep{Aftab_2025_arXiv} (DR1), which is well-matched to SPHEREx in terms of spatial resolution (Figure~\ref{fig:rosette_pa_halpha}, middle). In the right panel of Figure~\ref{fig:rosette_pa_halpha}, we show the logarithmic ratio of Pa$\alpha$ to H$\alpha$ relative to the median ratio over the Rosette nebula. The median ratio of 0.254 exceeds the Case B value ($\approx0.12$ at $10^4$\,K; \citealt{Hummer_1987_MNRAS, Storey_1995_MNRAS}) by an amount corresponding to $A_V \approx 1.3$\,mag for the \citet{Cardelli_1989_ApJ} extinction curve assuming $R_V = 3.1$, consistent with the $A_V \approx 1.2$--$1.5$\,mag of integrated foreground extinction to the distance of NGC 2244 in a GaiaXP-based 3D dust map \citep{Edenhofer_2024_AA}.

\begin{figure*}[t]
    \centering
    \includegraphics[width=0.85\linewidth]{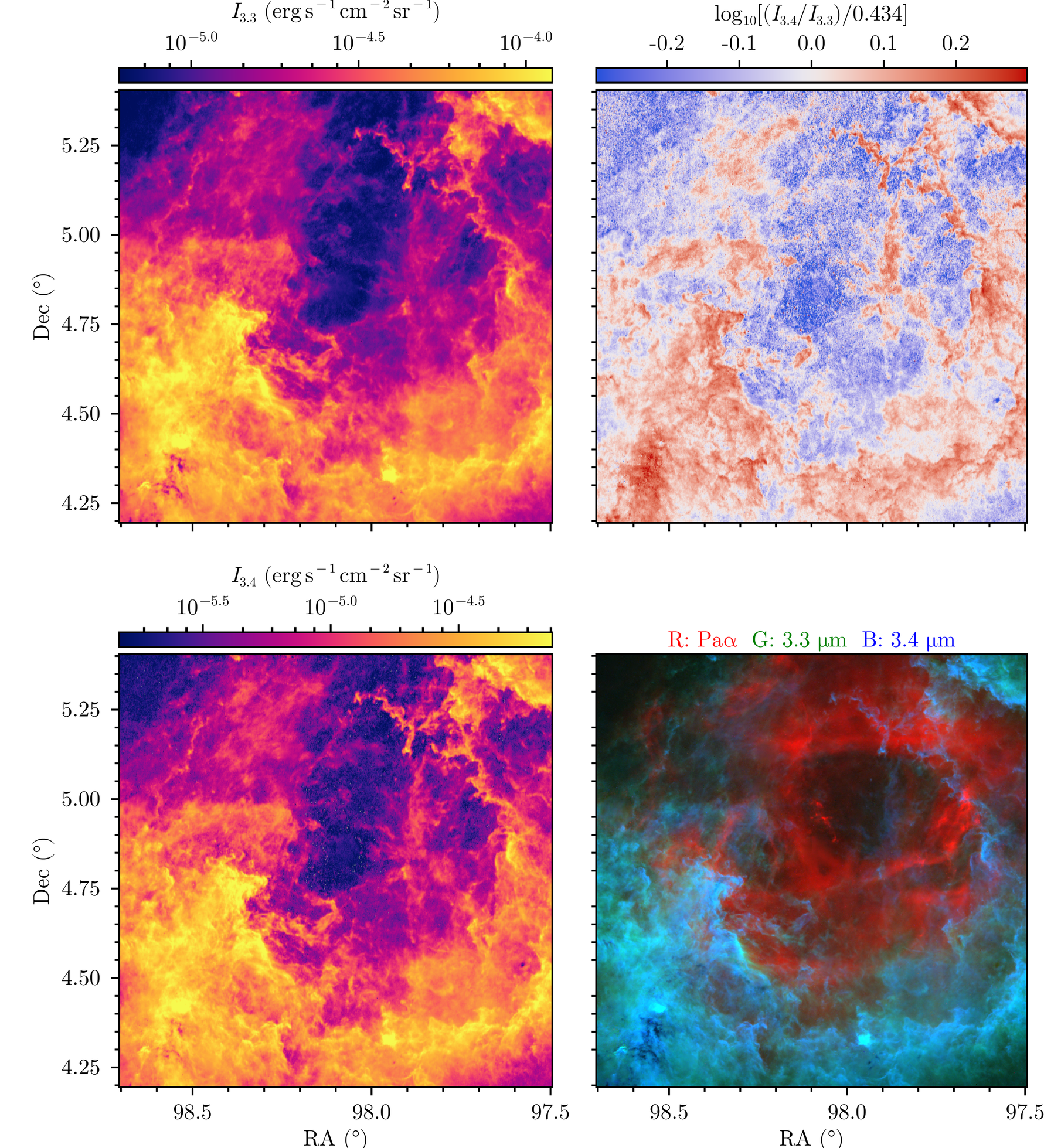}
    \caption{PAH 3.3\um (top left) and 3.4\um (bottom left) line maps for the Rosette nebula. The aliphatic-to-aromatic ratio $\log_{10}[(I_{3.4}/I_{3.3})/0.434]$ is shown in the top right, where 0.434 is the median ratio over the displayed region. Red regions indicate locations with enhanced 3.4\um compared to the average. Bottom right is an RGB composite (R: Pa$\alpha$, G: 3.3\um, B: 3.4\um, each normalized by its median and asinh-stretched) which shows how the ionized shell traced by Pa$\alpha$ lies inside the PAH-emitting photodissociation region.}
    \label{fig:rosette_4_3_ratio}
\end{figure*}

These SPHEREx PAH and Pa$\alpha$ maps thus provide complementary probes of the Rosette nebula. We expect future work will combine high resolution PAH and hydrogen recombination line maps from SPHEREx, high resolution ground-based H$\alpha$ maps from MDW, and lower-spatial resolution but much higher spectral resolution integral field unit (IFU) maps from the SDSS-V Local Volume Mapper (LVM) \citep{Villa-Durango_2025_MNRAS} in order to significantly progress efforts in reconstructing the 3D density, temperature, and ionization structure of the Rosette nebula.

\subsubsection{Aliphatic-to-Aromatic Ratio} \label{sec:ali2aro}

Because the 3.3\um PAH is primarily associated with aromatic carbon-hydrogen bond stretches and the 3.4\um PAH with aliphatic carbon-hydrogen stretches, the 3.4\um-to-3.3\um PAH ratio acts as a tracer of chemical variation. We display maps of the aliphatic-to-aromatic ratio for the Rosette nebula in Figure~\ref{fig:rosette_4_3_ratio}. The morphology of the ratio map is strikingly distinct from the morphology of the 3.3\um line map: the ratio map shows relatively enhanced 3.4\um emission in the smaller, clumpier structures that line the boundaries of the \ion{H}{2} region, and a relative dearth of 3.4\um emission both interior to the Rosette shell and in the more diffuse regions in the outskirts of the Rosette nebula. This is generally consistent with trends observed with SPHEREx data toward the Iris nebula \citep{Boersma_2026_ApJ}, but LVFit enables a spectacularly spatially resolved view of the relative strength of these two spectral features.
Interpreting the 3.4\um feature as arising primarily from aliphatic groups attached to PAHs \citep{Allamandola_1985_ApJL, Tielens_2008_ARAA}, we infer that these moieties or molecules are more easily photolyzed when exposed to far-UV photons, and thus only survive in relative abundance when shielded from the far-UV radiation field in denser clumps of dust \citep{Joblin_1996_ApJ}. This interpretation is consistent with the particularly strong aliphatic-to-aromatic ratio coincident with the Rosette molecular cloud and the dense edges of the northern shell. We plan to leverage the power of our enhanced spatial resolution in future work to carefully model the destruction thresholds for these PAHs in terms of incident UV flux and dust density.

While the absolute ratio of 3.4/3.3\um PAH emission depends strongly on convention for partitioning overlapping spectral features, including the continuum, we can more easily compare the relative values we measure here. Compared to the PDR shell ($0\fdg3$--$0\fdg6$), the median $I_{3.4}/I_{3.3}$ inside the \ion{H}{2} cavity ($r<0\fdg25$) is 15\% lower and within $0\fdg3$ of the Rosette molecular cloud central core is 14\% higher. We find 3.4/3.3\um ratios are $\sim15\%$ higher than the Rosette average in our molecular cloud testbeds such as Corona Australis (Section~\ref{sec:corAus}). Mapping and understanding the variations of this ratio in detail across all Galactic environments would be a valuable topic for future work.

Under our 3.4/3.3\um convention, the median absolute ratio in Rosette is 0.44. To compare to \citet{Mori_2014_ApJ}, who studied the 3.4/3.3\um ratio in 36 Galactic \ion{H}{2} regions, we estimate that assigning the PAH plateau beneath the 3.3 and 3.4\um PAH features to the continuum similar to their method would reduce our ratios by a factor of $0.42$, to $0.16$--$0.21$ for Rosette, which agrees well with their quoted ratios of $0.25\pm0.1$. These can then be converted into an upper bound on the fraction of C atoms in aliphatic chains versus aromatic rings, which gives $\sim3\%$ \citep{Yang_2016_ApJ}. This conversion leverages theoretical oscillator strengths to predict the 3.4/3.3\um ratio if there were equal numbers of aliphatic and aromatic C-H bonds, though the conversion factor varies by a factor of two depending on assumptions about what particular aliphatic species are present. We use the PAH continuum-subtracted ratio for this conversion because the PAH continuum in this region is often attributed to anharmonic hot bands \citep{Barker_1987_ApJL} and other transitions not included in the theoretical calculations of \citet{Yang_2016_ApJ}.

\begin{figure*}[t]
    \centering
    \includegraphics[width=\linewidth]{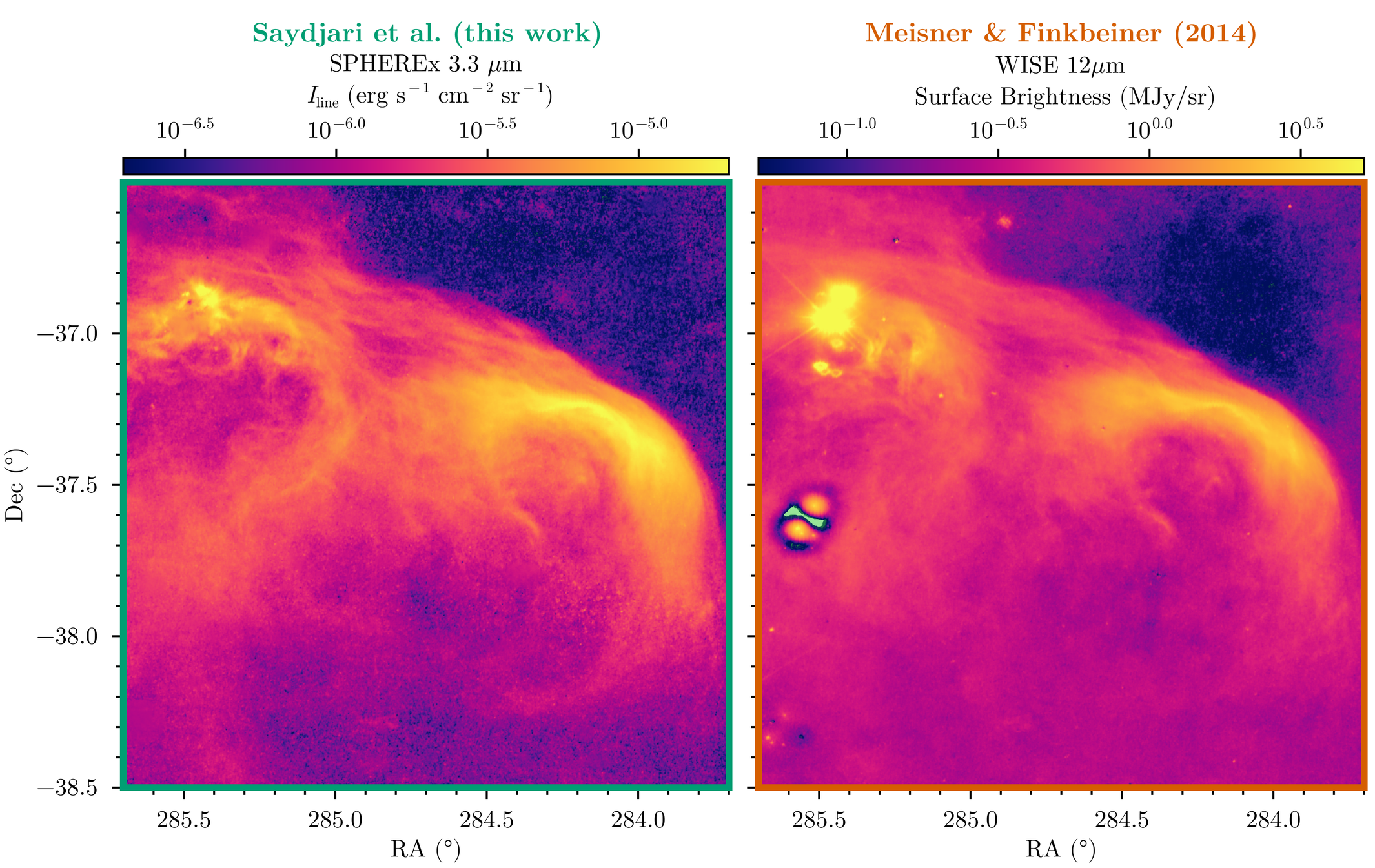}
    \caption{A portion of the Corona Australis dust cloud used to compare (left) the 3.3\um PAH map from LVFit and (right) the WISE 12\um PAH map \citep{Meisner_2014_ApJ}. To facilitate comparison, we smooth the LVFit map to match the $15\arcsec$ FWHM angular resolution of \citet{Meisner_2014_ApJ}.}
    \label{fig:corAus}
\end{figure*}

\subsection{Corona Australis} \label{sec:corAus}

Corona Australis is a nearby \citep[$\sim150$ pc,][]{Zucker_2020_AA} molecular cloud that is notable for its position well off the Galactic plane. Its proximity combined with its gradient in density makes it a great testbed for validating morphological recovery of an isolated cloud with variable signal-to-noise. We show the LVFit 3.3\um PAH map in the left panel of Figure~\ref{fig:corAus} and the 12\um WISE PAH map from \citet{Meisner_2014_ApJ} in the right panel, for comparison. While the WISE W3 PSF is $6\farcs5$ FWHM \citep{Wright_2010_AJ}, \citet{Meisner_2014_ApJ} smoothed their map to $15\arcsec$ FWHM and so we present the LVFit map smoothed to $15\arcsec$ FWHM as well to facilitate the comparison. 

In general, we find that the LVFit 3.3\um PAH map matches the 12\um WISE PAH map morphologically, but with fewer artifacts. The off-axis optical design of SPHEREx \citep{Bock_2026_ApJ} significantly reduces diffraction spikes compared to the on-axis design of WISE \citep{Wright_2010_AJ}, allowing us to more easily remove stars. This is especially apparent for the bright sources in the Coronet cluster at the top left of Figure~\ref{fig:corAus}, which also show significant pupil ghosts below the bright stars in the WISE map that are not present in our SPHEREx map. The WISE map also has a large defect at the middle left of the image associated with a satellite transit. On our first processing of the SPHEREx data, a faint linear streak was present in the bottom right of Figure~\ref{fig:corAus}, likely from a moving solar system object such as an asteroid. The affected measurements were subsequently excluded from their respective per-pixel fits by our streak detector (see Section~\ref{sec:contam}), which flags six such tracks in five of the 110 exposures of this field.

Another artifact in the WISE map results from \citet{Meisner_2014_ApJ} adjusting the zero-points to agree with Planck on scales above 2\degr to handle drifts in the WISE zero-point. This leads to ``ringing'' over the sharp edge of the cloud, where the 12\um emission appears to decrease and then return to a value typical of the rest of the low density regions in the image. In contrast, no such background adjustments were required for our SPHEREx processing and the background 3.3\um emission appears uniform in the top right of the image. We will discuss the sensitivity limits and comparisons in the low-density regime further in Section~\ref{sec:sens}. 

\begin{figure*}[p]
    \centering
    \includegraphics[width=\linewidth]{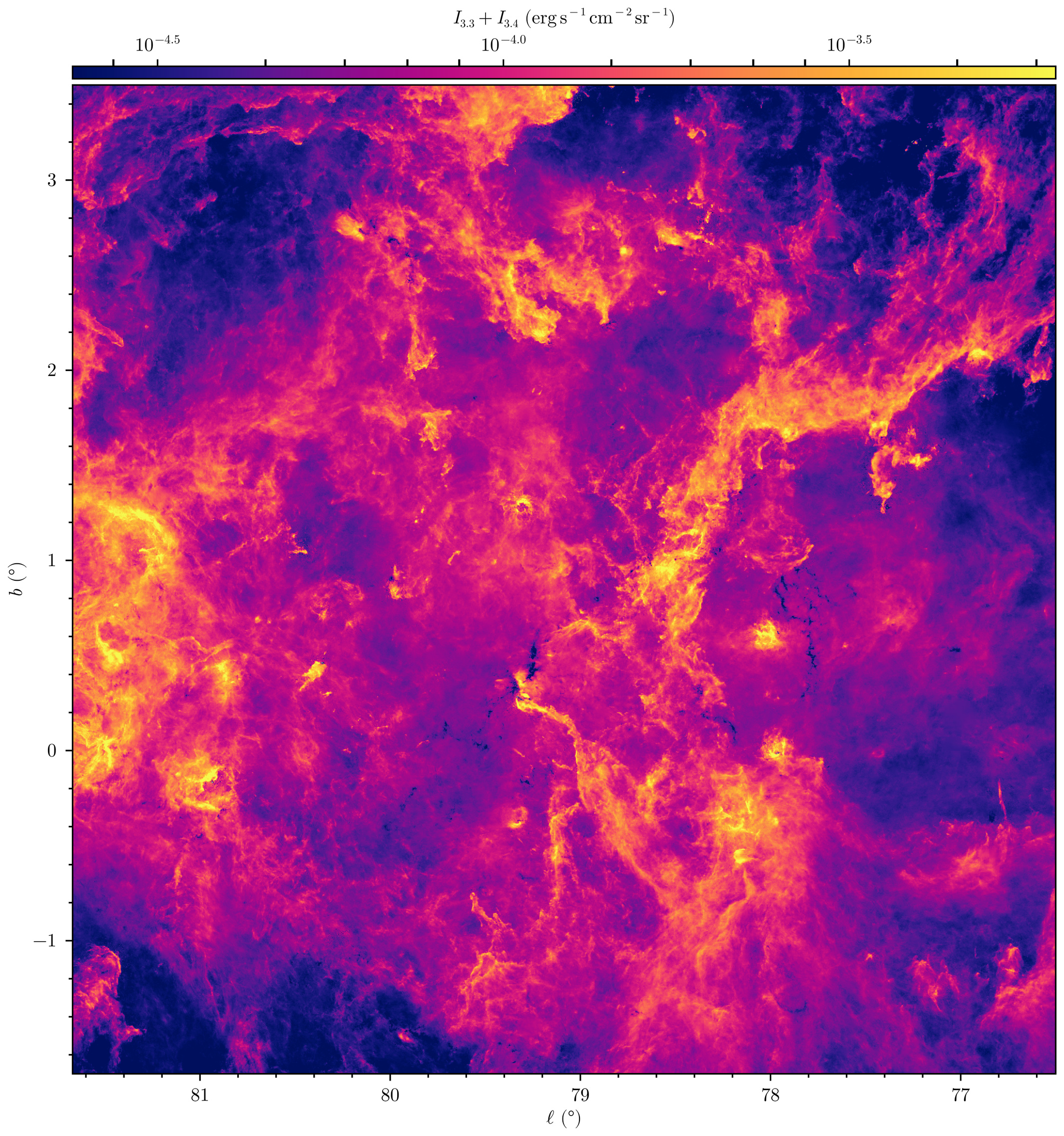}
    \caption{An LVFit map of Cygnus X in Galactic coordinates showing the sum of the 3.3 and 3.4\um PAH emission, $I_{3.3}+I_{3.4}$, on a log color scale spanning the 2$^{\rm nd}$--99.9$^{\rm th}$ percentiles. This field was studied by \citet{Hora_2026_ApJ}. About a third of the pixels in this crowded field are masked (mostly faint stars) and infilled.}
    \label{fig:cygX}
\end{figure*}

\begin{figure*}[p]
    \centering
    \includegraphics[width=\linewidth]{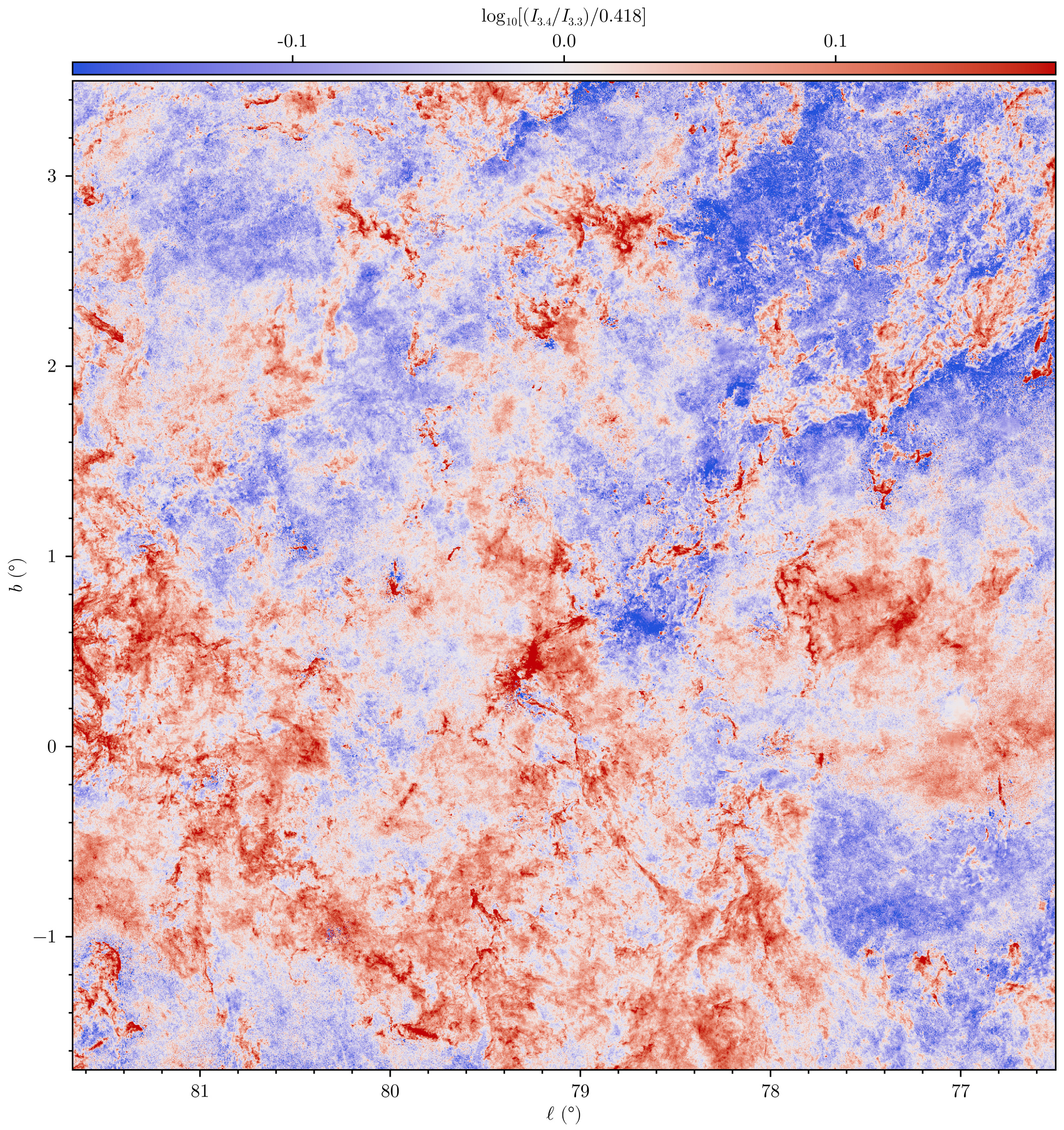}
    \caption{The aliphatic-to-aromatic ratio $\log_{10}[(I_{3.4}/I_{3.3})/0.418]$ for the Cygnus X field of Figure~\ref{fig:cygX}, where 0.418 is the median ratio over the displayed region. Red indicates enhanced 3.4\um emission relative to the median.}
    \label{fig:cygXratio}
\end{figure*}

\begin{figure*}[p]
    \centering
    \includegraphics[width=\linewidth]{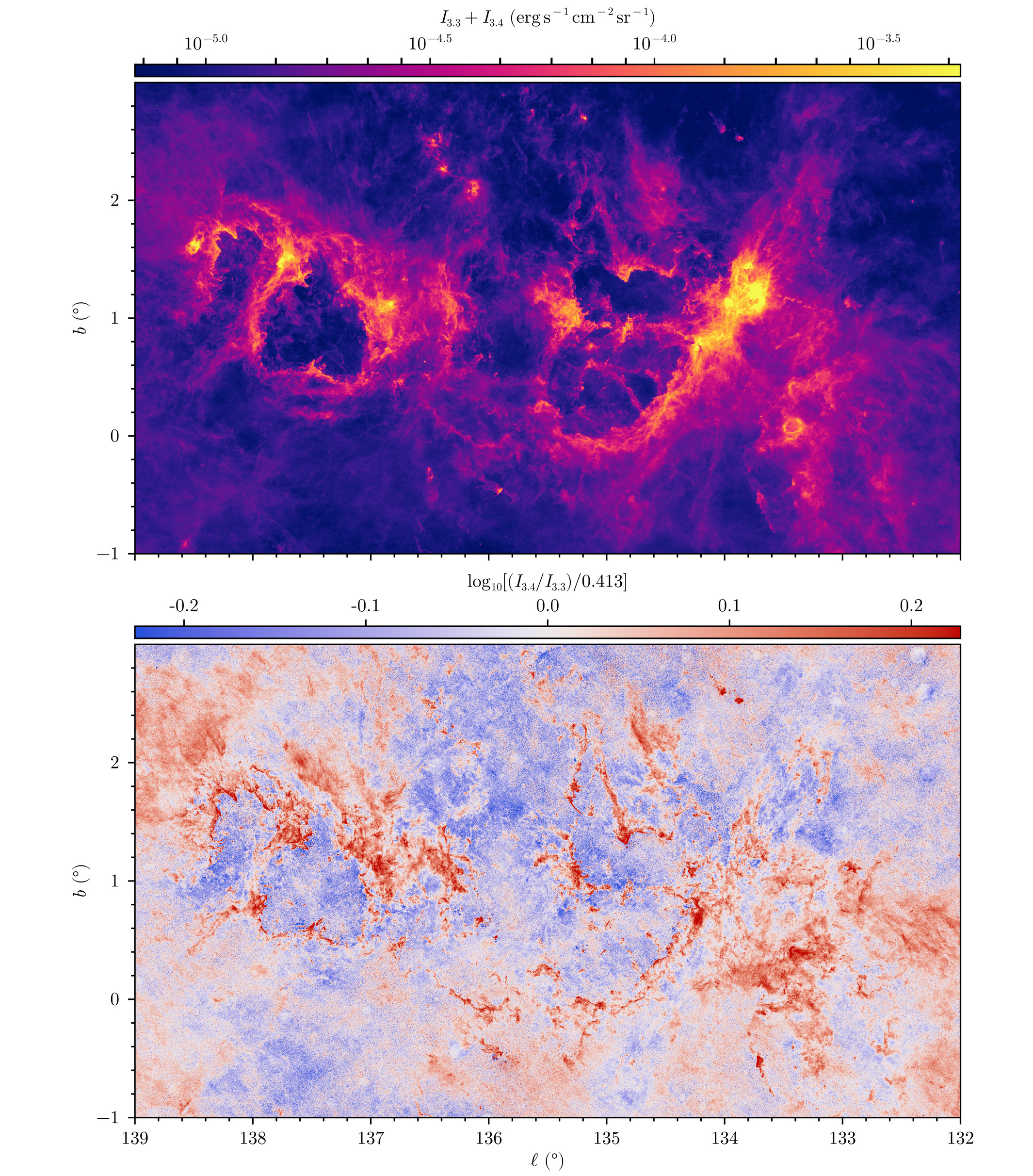}
    \caption{The Heart (IC 1805, W4) and Soul (IC 1848, W5) nebulae in Galactic coordinates, a $7\degr\times4\degr$ field centered on $(\ell, b) = (135\fdg5, +1\fdg0)$, which was studied by \citet{Cukierman_2026_ApJS}. Top: The sum of the LVFit 3.3 and 3.4\um PAH line maps, $I_{3.3}+I_{3.4}$, on a log color scale spanning the 2$^{\rm nd}$--99.9$^{\rm th}$ percentiles. Bottom: The aliphatic-to-aromatic ratio $\log_{10}[(I_{3.4}/I_{3.3})/0.413]$, where 0.413 is the median ratio over the displayed region. Red indicates enhanced 3.4\um emission relative to the median.}
    \label{fig:heartAndSoul}
\end{figure*}

After accounting for these artifacts, the ratio of the SPHEREx 3.3\um and the WISE 12\um maps could provide valuable insight into chemical variation between molecules traced by 3.3\um emission versus those traced by WISE W3. However, this comparison is far more fraught than interpreting the $3.4/3.3$\um maps because there are several different PAH emission complexes (7.7\um, 8.6\um, 11.3\um, 12.7\um) that all contribute to the WISE W3 band ($7.5$--$16.5$\um; \citealt{Wright_2010_AJ}). The relative strength of the 7.7\um and 11.3\um features within this band depends strongly on the PAH ionization state \citep{Draine_2021_ApJ}. 

These initial maps of Corona Australis are especially useful because the cloud will serve as an important testbed for at least two aspects of future work. First, literature ${\rm H}_2$ column density maps from Herschel \citep{Andre_2010_AA, HGBSTeam_2020_IRSA} will serve as an important reference for modeling ${\rm H}_2$ ro-vibrational transitions in SPHEREx. In addition, the reflection nebulae NGC 6726/6727/6729 near the Coronet cluster provide an important testbed for how to handle strongly spatially localized coupling between point sources and diffuse emission. When trying to make diffuse maps, the conundrum is always how to ensure extended light from point sources in the PSF wings is sufficiently removed while preserving any nearby diffuse structure. For the 3.3\um PAH map, this is even more pressing since the UV flux from stars may excite PAH emission in the nearby ISM, so there is a causal connection rather than simply a chance overlap between point sources and extended emission. Simple masking and infilling approaches like those described in this work, even after subtracting off our best model of the stellar light, often over-mask and miss some of the nearby substructure in these regimes. More advanced simultaneous modeling of point sources and the diffuse components may be required to better preserve the substructure of the reflection nebulae (see Section~\ref{sec:conc} for more).

\subsection{Cygnus X and the Heart and Soul Nebulae} \label{sec:miscOther}

We show two further Galactic testbed regions that have been featured in recent SPHEREx works. In Figures~\ref{fig:cygX} and \ref{fig:cygXratio} we show Cygnus X, which was studied by \citet{Hora_2026_ApJ}. Cygnus X hosts many embedded young stellar objects and compact \ion{H}{2} regions whose PAH emission our star mask partly removes, the same conundrum we discuss with regards to the Corona Australis reflection nebulae (Section~\ref{sec:corAus}). We plan to use this field to test the joint modeling of point sources and diffuse emission with spectral priors in order to better preserve these somewhat compact regions of diffuse emission. In Figure~\ref{fig:heartAndSoul}, we show the Heart and Soul nebulae, which were studied by \citet{Cukierman_2026_ApJS}. Neither makes the data behind their figures publicly available for comparison. 

\section{Extragalactic Testbeds} \label{sec:extragalactic}

\subsection{M31: Andromeda} \label{sec:m31}

\begin{figure*}[htb]
    \centering
    \includegraphics[width=0.835\linewidth]{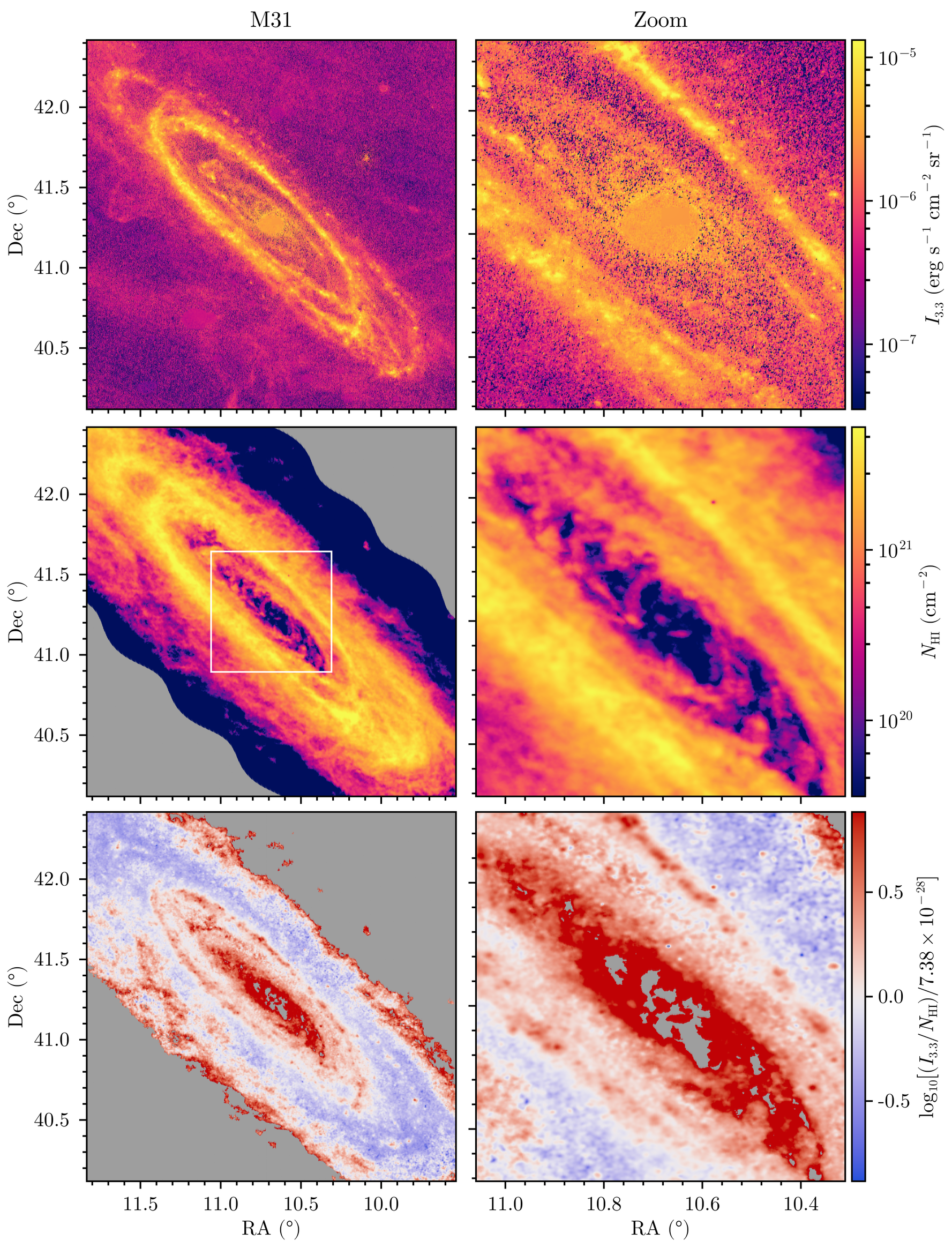}
    \caption{M31 (left) and a zoom-in (right) on the region marked by the box in the middle-left panel. Top: LVFit SPHEREx 3.3\um PAH line map. Middle: \ion{H}{1} column density from the Local Group L-Band Survey (LGLBS) VLA C+D+GBT moment-0 map \citep{Koch_2025_ApJS} at its native $24\arcsec$ beam, assuming optically thin emission. Gray pixels are areas outside the \ion{H}{1} mosaic. Bottom: $\log_{10}[(I_{3.3}/N_{\rm HI})/7.38\times10^{-28}]$, with our map convolved to $24\arcsec$ FWHM and restricted to \ion{H}{1} S/N$\,>3$, where we normalized by the average erg\,s$^{-1}$\,sr$^{-1}$ per H atom in the median outside the central $10\arcmin$. The PAH emission per H atom appears enhanced in the star-forming ring and toward the bulge, though starlight may contaminate the bulge measurement.}
    \label{fig:m31_pah}
\end{figure*}

In the top row of Figure~\ref{fig:m31_pah}, we show the $3.3$\um PAH map of M31 from LVFit, with a zoom-in on the center of M31 in the right column. Twelve moving-object tracks in 11 exposures have their pixels rejected by our streak detector, and 29 of the 253 exposures have the detector edge masked for scattered moonlight (see Section~\ref{sec:contam}). For comparison, we show the \ion{H}{1} column density at $24\arcsec$ from the Local Group L-Band Survey (LGLBS) \citep{Koch_2025_ApJS}. One important difference between these maps is that we integrate over the full line of sight, including foreground PAH emission from Galactic cirrus, while the LGLBS mosaic shown applies their broad ``signal mask'' to reject \ion{H}{1} not associated with M31. 

By comparing these maps, we see that our PAH map successfully resolves substructure within the rings of M31 and low density rings/spiral structure just outside the nucleus that are also traced by the \ion{H}{1} gas. However, in the nucleus, our map appears to show PAH emission. Due to the lack of \ion{H}{1} in that region in the LGLBS image and the low dust mass surface density measured by Spitzer \citep{Draine_2014_ApJ}, we interpret this as continuum emission from stars that is not fully removed in this extremely crowded limit. This same mechanism of leakage of continuum into the PAH map appears to impact the satellite dwarf elliptical galaxy M110 (top right quadrant of the left column), which is not expected to have significant PAH emission due to its low metallicity \citep{Goncalves_2014_MNRAS} near the threshold below which PAH emission is deficient \citep{Draine_2007_ApJ}, lack of ongoing star formation \citep{Monaco_2009_AA}, and low dust content \citep{DeLooze_2012_MNRAS}. Jointly forward modeling both the resolved and unresolved point sources with the diffuse emission is one possible solution to this limitation, which we discuss further in Section~\ref{sec:conc}. Revisiting M110 and similar objects after we have better control and quantification of continuum leakage may either provide a detection or valuable upper bound on PAH content in Local Group dwarf ellipticals.

In the bottom row of Figure~\ref{fig:m31_pah}, we compare the ratio of the PAH 3.3\um emission and \ion{H}{1} column density maps. In between the edges of the LGLBS map and the continuum contamination in the nucleus, we appear to clearly trace enhanced PAH emission in M31's ring known for enhanced star formation \citep{Gordon_2006_ApJL, Draine_2014_ApJ, Lewis_2015_ApJ}. A deeper investigation of individual sites of star formation and variation in the 3.3\um and 3.4\um content of those regions as a function of galactic environment would be an exciting avenue for future work.

\begin{figure*}[p]
    \centering
    \includegraphics[width=0.72\linewidth]{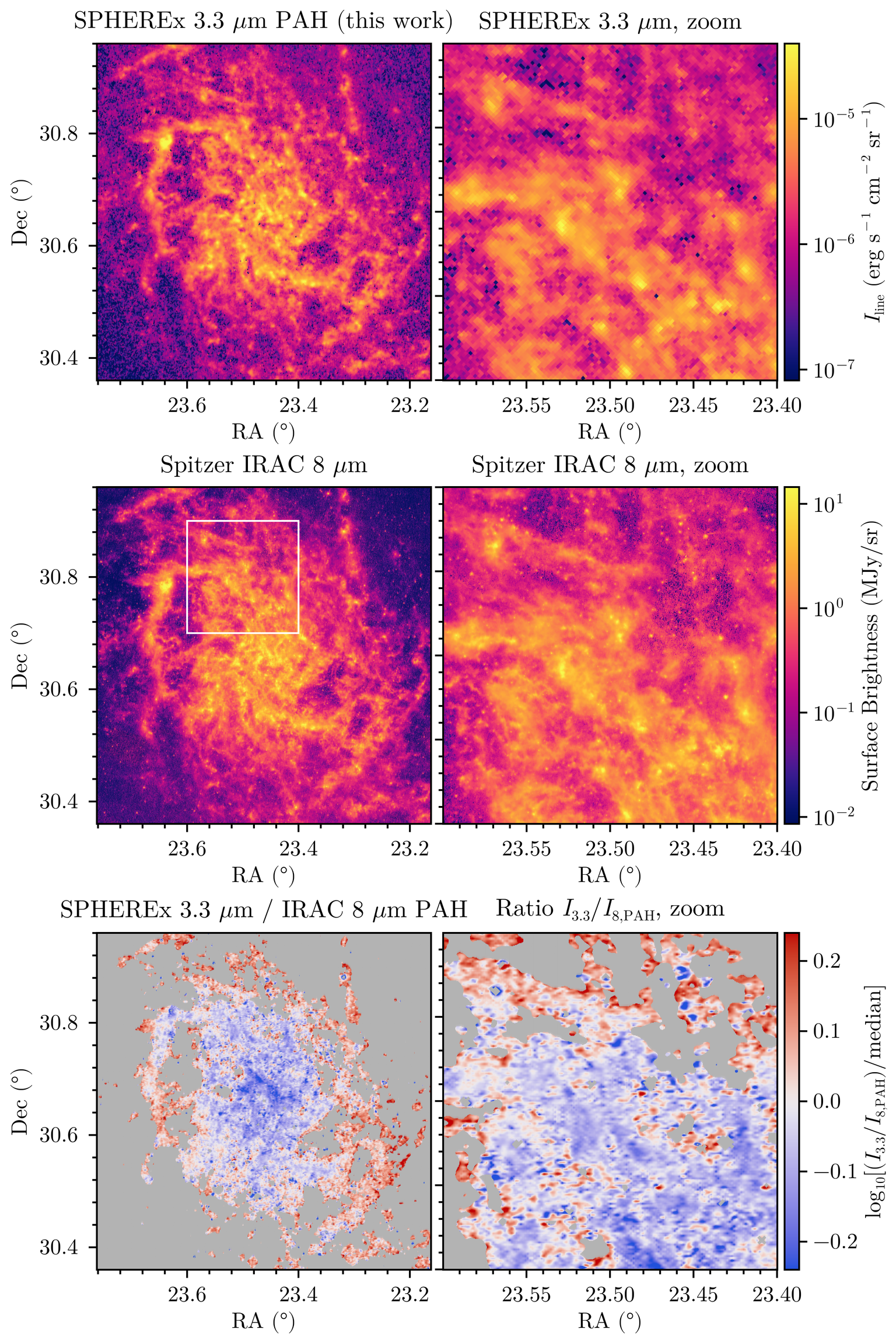}
    \caption{Top: The LVFit SPHEREx 3.3\um PAH map of M33 (left) and a zoom-in on the region marked by the white box in the middle left panel (right). Middle: The same regions in the Spitzer IRAC 8\um band, which is dominated by the 7.7\um PAH complex and is shown at its native $\sim2\arcsec$ FWHM angular resolution without smoothing, as a higher-resolution reference for the structure in the LVFit map ($8\farcs5$ FWHM). Bottom: The ratio of the LVFit 3.3\um map to the stellar-continuum-corrected 8\um PAH emission, $\log_{10}[(I_{3.3}/I_{8,\mathrm{PAH}})/\mathrm{median}]$, where both maps are interpolated and smoothed to a common $10\farcs7$ FWHM angular resolution. The median ratio is $4.3\times10^{-6}$\,erg\,s$^{-1}$\,cm$^{-2}$\,sr$^{-1}$ per MJy\,sr$^{-1}$. Pixels shaded gray are either in our star mask or have $I_{8,\mathrm{PAH}} < 0.25$\,MJy\,sr$^{-1}$.}
    \label{fig:m33_pah}
\end{figure*}

\subsection{M33: Triangulum} \label{sec:m33}

Similar to M31, the LVFit 3.3\um PAH map of M33 (Figure~\ref{fig:m33_pah}, top row) can be used to make maps of PAH-to-neutral-H and trace star formation in a Local Group galaxy. The LVFit map can also be compared to the simple continuum-subtracted SPHEREx 3.3\um M33 map presented in \citet{Cukierman_2026_ApJS}. As noted in \citet{Cukierman_2026_ApJS}, the SPHEREx data enable analyses of the ratio between the 3.3\um emission and the point-source-subtracted WISE 12\um map of M33 presented in \citet{Jarrett_2019_ApJS}, similar to our discussion in Section~\ref{sec:corAus}. 

One other important legacy dataset for comparison is Spitzer. As we show in the middle row of Figure~\ref{fig:m33_pah}, M33 has 8\um IRAC imaging from Spitzer \citep{Werner_2004_ApJS, Fazio_2004_ApJS} that looks extremely similar to the LVFit map, but which has a PSF with $2\arcsec$ FWHM. Here we show the sky level-subtracted Spitzer Enhanced Imaging Products super mosaic \citep[SEIP;][]{SEIP_2020_IRSA}. To put the mosaic tiles on a common background before forming the ratio, we solve for a constant offset per tile from the tile overlaps, which differ by up to a few tenths of a MJy sr$^{-1}$ at 8\um. One important difference from the LVFit SPHEREx map is that the 8\um IRAC image is not point source subtracted. In terms of PAH emission, the 8\um IRAC band covers 6.4--9.4\um and primarily traces the 7.7\um PAH feature, with contributions from 6.2\um and 8.6\um PAH features, all of which are typically associated with ionized PAH carriers (Table~\ref{tab:pah_features}). In contrast, the 3.3\um PAH is typically associated with aromatic C-H stretches in small neutral PAHs \citep{Draine_Li_2007, Maragkoudakis_2020_MNRAS, Draine_2021_ApJ}, and so careful analysis of the $3.3/7.7$\um ratio can trace variation in the charge state of PAHs in the ISM.

In the bottom row of Figure~\ref{fig:m33_pah} we show the $3.3/7.7$\um PAH ratio after interpolating and smoothing both maps to a common $10\farcs7$ FWHM angular resolution. We apply the stellar continuum correction of \citet{Helou_2004_ApJS} to obtain $I_{8,\mathrm{PAH}} = I_{8} - 0.232\,I_{3.6}$, which removes the stellar continuum using IRAC 3.6\um. This should be a good first order correction for variation in the stellar continuum, but the IRAC 3.6\um band used for correction does also contain 3.3\um PAH emission that may be important to model to measure the $3.3/7.7$\um PAH ratio in detail. Many of the bright knots in the ratio map are simply due to the lack of point-source subtraction in the 8\um IRAC image.

In the LVFit ratio map, we see that the $3.3/8$\um PAH ratio increases radially by $+0.048\pm0.003$ dex/kpc from the center to the edge of M33. The quoted uncertainty is statistical. The level of the 8\um background is the dominant systematic, and shifting it by the scatter of the mosaic tile offsets moves the slope by $\sim0.01$ dex/kpc. This gradient could be reflective of either variations in metallicity or the radiation field spectrum. Recently, \citet{Koziol_2026_arXiv} studied this question with JWST NIRCam imaging of 19 PHANGS galaxies and the PAH emission model grids from \citet{Draine_Li_2007} and \citet{Draine_2021_ApJ} to conclude the 3.3\um PAH radial gradient is driven primarily by metallicity, with a slope of $\mathrm{d}\log_{10}(I_{3.3}/I_{7.7})/\mathrm{d}(12+\log_{10}(\mathrm{O/H})) = -0.85 \pm 0.08$. We can compare our observations to this using the metallicity gradient measured by the CHemical Abundances Of Spirals (CHAOS) project from nebular and auroral emission lines in 65 \ion{H}{2} regions with the Large Binocular Telescope (LBT) \citep{Rogers_2022_ApJ}. They find $-0.037\pm0.007$ dex/kpc in $12+\log_{10}(\mathrm{O/H})$. Combined, this gives us $\mathrm{d}\log_{10}(I_{3.3}/I_{8})/\mathrm{d}(12+\log_{10}(\mathrm{O/H})) = -1.3 \pm 0.3$, which is consistent with \citet{Koziol_2026_arXiv} to within $2\sigma$. We speculate that some differences could come from comparing a single galaxy to a trend pooled over many, and from our use of the 3.3\um feature alone, since their 3.3\um band also contains the 3.4\um complex. Adding our 3.4\um map to the numerator does flatten the slope, because the 3.4/3.3\um ratio itself declines outward in M33, though we leave a more careful exploration of this to future work.

While these specific results are mostly validations of recent JWST results, the spectral information in LVF imaging from SPHEREx can enable better component separation (e.g. continuum subtraction, 3.3\um versus 3.4\um emission) compared to imaging with classical photometric filters, which may allow a resolution to open questions about the mechanism(s) driving the radial gradient. In addition, SPHEREx provides access to these 3.3\um and 3.4\um PAH emission maps all-sky without expensive targeted imaging/spectroscopy, opening up a wide variety of targets for study.

Beyond these astrophysical motivations for studying M33, the higher resolution and sensitivity of the 8\um IRAC data make M33 a perfect testbed for (1) attempts at super-resolution, by forward modeling sub-PSF scale shifts between different exposures, (2) validating models of undetected point sources, and (3) validating differentiation of compact diffuse emission from point sources. This will be the subject of future work (see Section~\ref{sec:conc}).

\section{Discussion} \label{sec:disc}

\subsection{Sensitivity} \label{sec:sens}

One key property of a diffuse emission map is the sensitivity it achieves. We prefer an empirical estimate of this precision measured on the LVFit maps in order to compensate for any limitations in our uncertainty model, which we discuss further in Section~\ref{sec:conc}. We use this sensitivity analysis to calibrate the LVFit per-pixel uncertainty maps delivered with the data products (Section~\ref{sec:dataavil}).

We have to be careful about which pixels we select for an empirical sensitivity estimate, because each \healpix $N_{\rm side} = 2^{15}$ pixel is fit with the $6\farcs44$ FWHM Gaussian spatial weighting truncated at $5\farcs5$, as described in Section~\ref{sec:linfit}. As a result, only pixels that are at least two pixels apart use a disjoint set of measurements. Assuming that the structure in our image varies slowly relative to the pixel scale, we can then measure the scatter between pairs of pixels separated by two pixels as an empirical estimate of the pixel-to-pixel uncertainty in the map. To account to first order for spatial variation in the map, we actually compute $d_r = I_r - (I_{r-2} + I_{r+2})/2$ for pixels along the \healpix ring ordering index $r$ and estimate the single pixel noise at index $r$ as $\sigma_r = 1.4826\,\rm{MAD}(d_r)/\sqrt{1.5}$, where we use the median absolute deviation as an outlier-robust measure of spread, and convert to $\sigma_r$ using its relation to $\sigma$ for a unit normal distribution. 

These per-pixel estimates can then be used to estimate uncertainties for averages on large spatial scales by roughly scaling by $\sqrt{N}$ for an average over $N$ pixels. The estimator $\sigma_r$ is noisy and so we average the $\sigma_r$ over the parent \healpix $N_{\rm side} = 2^{11}$ ($\sim1\farcm7$) pixels, four levels up from the map grid. This gives approximately 230 samples per $N_{\rm side} = 2^{11}$ pixel, which we find yields an estimator of the noise with an uncertainty of $\sim10\%$. This provides sufficient precision for us to fit analytic models to our noise estimates and assess the quality of those fits.

This estimate of the precision is conservative in the sense that astrophysical structure causing variation between pixels beyond a linear gradient will inflate our estimate. On the other hand, systematics in common between different detector pixels in the same exposure may be shared between neighboring pixels. This is one way in which the pixels we use are not completely independent, even though they are at least two pixels apart. However, given that most pixels have $\gtrsim35$ exposures, we expect these systematics to predominantly enter as additional scatter between pixels covered by different combinations of exposures. Given the highly structured nature of Galactic emission on all measured scales, we expect true astrophysical structure to dominate the measured variation, thereby making our precision estimate conservative.

\begin{figure*}[t]
    \centering
    \includegraphics[width=\linewidth]{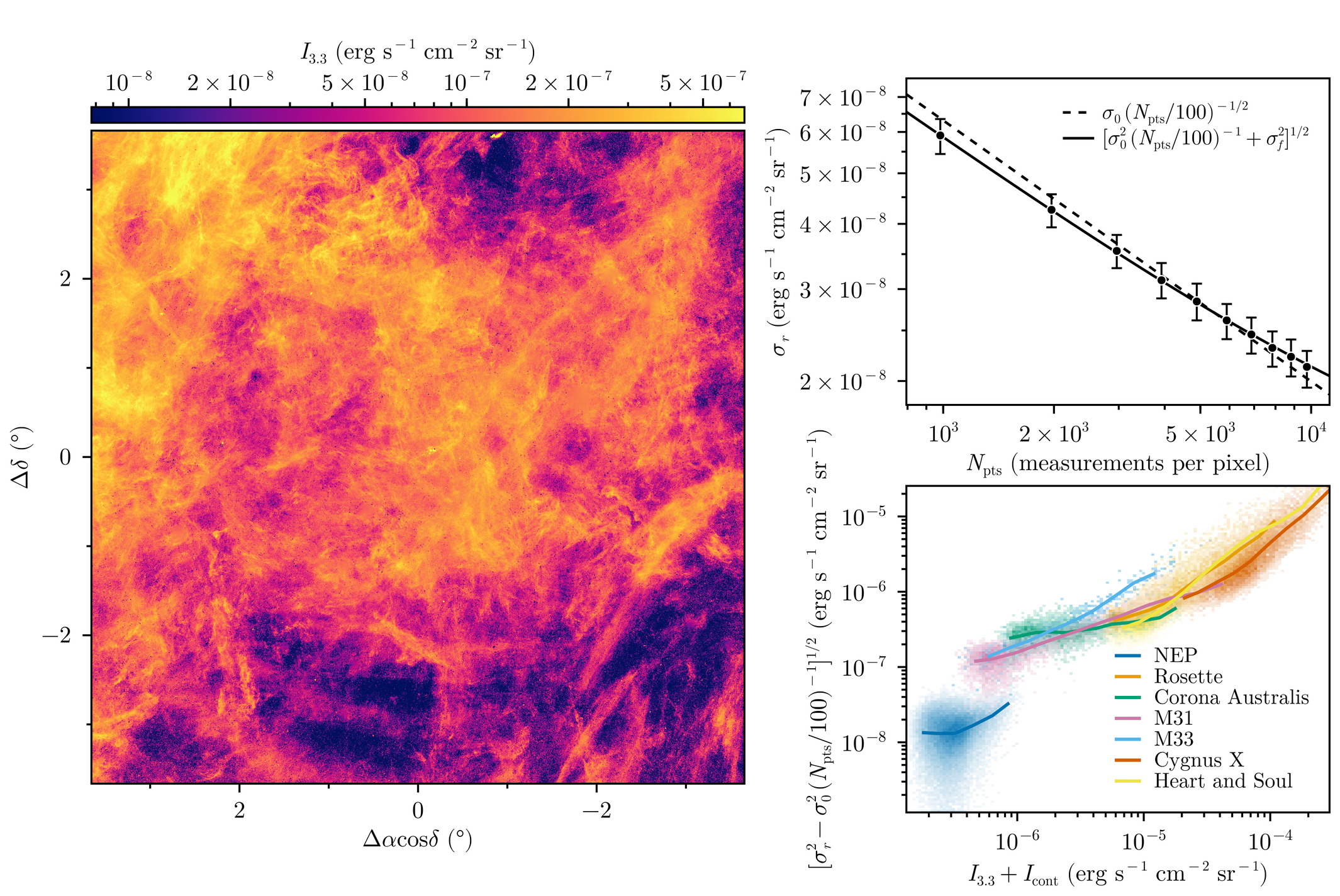}
    \caption{Left: A $7\fdg3\times7\fdg3$ field of the LVFit 3.3\um PAH map centered on the north ecliptic pole in gnomonic offsets from (RA, Dec) = ($270\fdg0$, $+66\fdg56$) with east to the left. Top right: Average pixel noise $\sigma_r$ in the central $1\degr\times1\degr$ of this field versus the median number of measurements per pixel $N_{\rm pts}$, while varying the number of exposures kept. The error bars show the interquartile range of $\sigma_r$ averaged over parent $N_{\rm side} = 2^{11}$ pixels. The dashed line is pure $N_{\rm pts}^{-1/2}$ scaling and the solid line adds a floor in quadrature. Bottom right: The excess noise above the $N_{\rm pts}^{-1/2}$ scaling, using $\sigma_0$ from the solid-line fit versus the total 3.3\um surface brightness $I_{\rm tot}$ in the fields shown in this work. The shading shows density of samples for each field (normalized by peak density of that field) and lines are running medians.}
    \label{fig:nep_prec}
\end{figure*}

We want to establish how the per-pixel precision scales with the number of measurements used in our fits. For this, we will consider the NEP (Figure~\ref{fig:nep_prec}, left), one of the two deep fields where SPHEREx's observational strategy accumulates the largest number of exposures. We can compute $\sigma_r$ across the field, dropping a varying number of exposures, keeping between 10\% and 100\% of the exposures in steps of 10\% (Figure~\ref{fig:nep_prec}, top right). From this we can estimate that our $\sigma_r = \sigma_0\times (N_{\rm pts}/100)^{-1/2}$ with $\sigma_0 = 2.00 \times 10^{-7}$ erg s$^{-1}$ cm$^{-2}$ sr$^{-1}$. However, we find that the scaling with $N_{\rm pts}$ is better fit with an uncertainty floor added in quadrature, for which we find $\sigma_0 = 1.82 \times 10^{-7}$ and $\sigma_f = 1.11 \times 10^{-8}$ erg s$^{-1}$ cm$^{-2}$ sr$^{-1}$.

There are several possible origins of this apparent floor. The most likely stems from multiplicative errors in the forward model, such as in the component shapes or the spectral response of the LVF. To evaluate this, we can plot $\sqrt{\sigma_r^2 - \sigma_0^2\times (N_{\rm pts}/100)^{-1}}$ versus the total surface brightness at 3.3\um for a variety of our test fields (Figure~\ref{fig:nep_prec}, bottom right). This suggests it is necessary to add a term in quadrature that is a function of surface brightness. However, the fact that not all of the fields fall along the same line suggests that the multiplicative term is different for the different fields. Specifically, the NEP has much higher precision than the other fields. Understanding the origin of these differences, for example whether the NEP benefits from seeing a wider range of detector orientations or if astrophysical differences between regions drives variations in model-data mismatches, will be left to future work. For now, we fit the multiplicative term for each field (Table~\ref{tab:noise}), adopting
\begin{equation} \label{eq:noise_sigma0}
    \sigma_r^2 = \sigma_0^2 \left(\frac{N_{\rm pts}}{100}\right)^{-1} + \left(f\, I_{\rm tot}\right)^2,
\end{equation}
with $\sigma_0$ and $f$ both fit per field and $I_{\rm tot} = I_{3.3} + I_{\rm cont}$ the total surface brightness through the 3.3\um feature, where $I_{\rm cont} = c_{3.29}\,\Delta\nu$ is the continuum surface brightness fit at 3.29\um integrated over the feature width $\Delta\nu = c\sqrt{2\pi}\,\sigma_\lambda/\lambda^2$ ($\sigma_\lambda = 0.017$\um, so 1 MJy sr$^{-1}$ of continuum corresponds to $1.19\times10^{-5}$ erg s$^{-1}$ cm$^{-2}$ sr$^{-1}$). The fit $\sigma_0$ is $1.9\times10^{-7}$ in the NEP, consistent with the depth test above, and $2.3$--$3.1\times10^{-7}$ in the other fields, as expected from their brighter zodiacal background at lower ecliptic latitude. The fit values for $f$ range from 0.03 (NEP) to 0.14 (M33). Away from the ecliptic poles, $N_{\rm pts}$ varies by less than a factor of 2 within a field, so $\sigma_0$ there is set by the noise of the faintest patches rather than by the $N_{\rm pts}$ dependence, which we only test in the NEP. In Cygnus X every patch is bright enough that the surface-brightness term dominates and $\sigma_0$ is unconstrained.

To convert from $N_{\rm pts}$ used above to the number of exposures, we note that the number of points-per-exposure passing our wavelength and spatial cuts of our PAH model is approximately 2.3. With the current 16.5 months of SPHEREx imaging, pixels in regions at low ecliptic latitude typically have 35 exposures ($N_{\rm pts} \approx 80$), while the ecliptic poles can have two orders of magnitude more.

We can also compare our formal LVFit uncertainties $\sigma_{\rm formal}$ (from the Huber-weighted covariance of the fit) with the empirical $\sigma_r$ measurements. We recalibrate the per-pixel uncertainties for each field with the same functional form,
\begin{equation} \label{eq:noise_formal}
    \sigma_r^2 = \left(k\, \sigma_{\rm formal}\right)^2 + \left(f'\, I_{\rm tot}\right)^2,
\end{equation}
fitting $k$ and $f'$ on the same $1\farcm7$ patches (Table~\ref{tab:noise}). Because $\sigma_{\rm formal}$ tracks the number, variances, and weights of the measurements that enter each pixel, it already carries most of the depth and some of the brightness dependence. A single scale factor $k$ and a small surface-brightness term then describe $\sigma_r$ with the same or lower scatter than Equation~\ref{eq:noise_sigma0} in every field. In the NEP, Corona Australis, M31, and M33 the formal uncertainties are within 5--25\% of the empirical noise ($k = 0.76$--$0.94$) and the recalibrated uncertainties reproduce $\sigma_r$ with a scatter of 0.05--0.06 dex over patches, which is within the precision of our $\sigma_r$ estimator. In the bright Galactic plane fields the formal uncertainties underestimate the scatter by a factor of 1.2--2.6 at low surface brightness, and the surface-brightness term dominates. In Cygnus X, the best fit sets $k = 0$ and the noise is simply $0.039\,I_{\rm tot}$ (as in Equation~\ref{eq:noise_sigma0}), with a scatter of 0.17 dex. A single global recalibration ($k = 0.88$, $f' = 0.038$) reproduces the NEP, Corona Australis, and M31 to within 6\% but is biased by 0.1 dex for M33, so we report the per-field values and deliver the recalibrated per-pixel uncertainties with our data products. The same analysis of the tied hydrogen recombination-line amplitude (Section~\ref{sec:methods}) gives formal uncertainties that are already within 20\% of the empirical noise ($k = 1.0$--$1.2$), a surface-brightness term is not required there, and the fit $f'$ values are consistent with zero. For the Pa$\alpha$ map of the Rosette nebula (Section~\ref{sec:paa}, fit on a separate detector) the formal uncertainties are within 10\% of the noise at low surface brightness and the empirical noise again requires a surface-brightness term ($k = 1.08$, $f' = 0.024$), with a fraction similar to that of the 3.3\um maps.

\providecommand{\grouphead}{}\renewcommand{\grouphead}[1]{\noalign{\vskip 1.2ex}\hline\noalign{\vskip 0.9ex}\multicolumn{9}{c}{\parbox{0.9\textwidth}{#1}}\\ \noalign{\vskip 0.9ex}\hline\noalign{\vskip 0.4ex}}
\begin{deluxetable*}{l|cc|ccc|ccc}
\tabletypesize{\small\tabcolsep=9.5pt}
\tablecolumns{9}
\tablewidth{0pt}
\tablecaption{Per-field noise models. Rows are grouped by field and quantity being mapped, with the basis vectors of the corresponding fit listed in the group header. $\sigma_r$ is the median empirical pixel noise and $N_{\rm pts}$ the median number of measurements per pixel. We show coefficients and residual scatter for fits to Equation~\ref{eq:noise_sigma0} and Equation~\ref{eq:noise_formal}. $\sigma_r$ and $\sigma_0$ are in units of $10^{-7}$ erg s$^{-1}$ cm$^{-2}$ sr$^{-1}$; $f$, $f'$, and $k$ are dimensionless. Values of $f$, $f'$ below 0.01 are consistent with zero and values of residual scatter below 0.05 dex are consistent with our $\sigma_r$ estimator uncertainty.\label{tab:noise}}
\tablehead{\multicolumn{1}{c|}{} & \multicolumn{2}{c|}{} & \multicolumn{3}{c|}{Equation~\ref{eq:noise_sigma0}} & \multicolumn{3}{c}{Equation~\ref{eq:noise_formal}} \\ \cline{4-6} \cline{7-9} \multicolumn{1}{c|}{Field} & \colhead{\makebox[32pt]{$\sigma_r$}} & \multicolumn{1}{c|}{\makebox[32pt]{$N_{\rm pts}$}} & \colhead{\makebox[32pt]{$\sigma_0$}} & \colhead{\makebox[32pt]{$f$}} & \multicolumn{1}{c|}{\makebox[32pt]{scatter}} & \colhead{\makebox[32pt]{$k$}} & \colhead{\makebox[32pt]{$f'$}} & \colhead{\makebox[32pt]{scatter}} \\ \multicolumn{1}{c|}{} & \colhead{($10^{-7}$)} & \multicolumn{1}{c|}{(median)} & \colhead{($10^{-7}$)} & \colhead{} & \multicolumn{1}{c|}{(dex)} & \colhead{} & \colhead{} & \colhead{(dex)}}
\startdata
\grouphead{\textbf{3.3\um PAH map:} 3.3+3.4\um PAH + continuum\tablenotemark{a}}
NEP & 0.27 & 5515 & 1.92 & 0.029 & 0.053 & 0.871 & 0.029 & 0.049 \\
Rosette & 8.37 & 80 & 2.92 & 0.052 & 0.130 & 0.716 & 0.049 & 0.126 \\
Corona Australis & 3.54 & 82 & 3.09 & 0.036 & 0.066 & 0.944 & 0.018 & 0.051 \\
M31 & 2.32 & 120 & 2.32 & 0.073 & 0.066 & 0.848 & 0.045 & 0.047 \\
M33 & 3.82 & 150 & 2.36 & 0.141 & 0.061 & 0.761 & 0.133 & 0.061 \\
Cygnus X & 23.5 & 118 & $<3$\tablenotemark{b} & 0.039 & 0.169 & $<0.01$ & 0.039 & 0.169 \\
Heart and Soul & 5.04 & 129 & 2.34 & 0.043 & 0.104 & 0.689 & 0.039 & 0.100 \\
\grouphead{\textbf{Pf$\delta$ map:} 3.3+3.4\um PAH + continuum + Brackett/Pfund/Humphreys lines + nebular continuum}
Rosette & 1.01 & 80 & 0.83 & $<0.01$ & 0.106 & 1.008 & $<0.01$ & 0.057 \\
Heart and Soul & 0.78 & 129 & 0.85 & $<0.01$ & 0.088 & 1.179 & $<0.01$ & 0.061 \\
\grouphead{\textbf{Pa$\alpha$ map:} Pa$\alpha$/Br$\delta$/Br$\epsilon$ + continuum}
Rosette & 19.0 & 44 & 6.79 & 0.030 & 0.115 & 1.080 & 0.024 & 0.103 \\
\enddata
\tablenotetext{a}{+ hydrogen recombination template for Rosette and Heart and Soul.}
\tablenotetext{b}{The noise is flux-dominated throughout the field, and any $\sigma_0 \lesssim 3\times10^{-7}$ erg s$^{-1}$ cm$^{-2}$ sr$^{-1}$ fits equally well.}
\end{deluxetable*}

An important point of comparison is the depth of this SPHEREx PAH map compared to the WISE 12\um PAH map from \citet{Meisner_2014_ApJ}. Trying to achieve comparable sensitivity appears difficult from the outset because the PAH emission in WISE W3 compared to the 3.3\um PAH feature is $\sim15:1$ in band-integrated power for standard diffuse-ISM PAH models \citep{Draine_2007_ApJ, Draine_2021_ApJ}, where W3/$I_{3.3} \approx 6 \times 10^4$ MJy sr$^{-1}$ per erg s$^{-1}$ cm$^{-2}$ sr$^{-1}$, and is larger still for PAH size distributions shifted to larger grains or for softer starlight spectra ($1$--$3 \times 10^5$ from the \citet{Draine_2021_Dataverse} model library). 

To convert the units between these two maps, we use the ratio actually measured on the test fields studied here, W3/$I_{3.3} = 1.8 \times 10^5$ MJy sr$^{-1}$ per erg s$^{-1}$ cm$^{-2}$ sr$^{-1}$ (measured as a robust slope, with a field-to-field scatter of 0.1 dex), which lies within the model range for larger-than-standard PAHs. These were also the models preferred by JWST observations in nearby galaxies \citep{Sandstrom_2023_ApJL}. The slope fit is lowest in the Galactic plane fields and highest in the high-latitude cirrus, the direction expected for harder versus softer radiation fields, which could be explored more in combination with recent 3D interstellar radiation field maps \citep{Lomeli_2026_arXiv, Bish_2026_arXiv} in future work. We then measured the empirical scatter as above on our PAH map, but smoothed to the $15\arcsec$ FWHM angular resolution of \citet{Meisner_2014_ApJ}, with both the WISE artifact mask and our infilling mask applied to both maps.

In the NEP, the WISE W3 precision is $3.6\times10^{-3}$ MJy sr$^{-1}$ corresponding to $1.7\times10^{-8}$ erg s$^{-1}$ cm$^{-2}$ sr$^{-1}$ in $I_{3.3}$. Converting the LVFit precision to the same spatial scale, we have $1.3\times10^{-8}$ erg s$^{-1}$ cm$^{-2}$ sr$^{-1}$ in $I_{3.3}$, which means the LVFit SPHEREx map is already a deeper PAH tracer at the ecliptic poles by a factor of 1.3. For the fainter fields away from the ecliptic poles, the W3 precisions are $8$--$15\times10^{-3}$ MJy sr$^{-1}$ corresponding to $4.5$--$8.0\times10^{-8}$ erg s$^{-1}$ cm$^{-2}$ sr$^{-1}$ in $I_{3.3}$, compared to the LVFit range in the same fields at the WISE spatial scale of $7.0$--$10\times10^{-8}$ erg s$^{-1}$ cm$^{-2}$ sr$^{-1}$ in $I_{3.3}$. Because the polar scan strategy of WISE \citep{Wright_2010_AJ} makes its coverage grow smoothly away from the ecliptic (from 10 exposures at $|\beta| < 20\degr$ to 25 at $|\beta| = 30$--$60\degr$ and 220 at the pole based on the \citet{Meisner_2014_ApJ} coverage maps), whereas the SPHEREx coverage only increases above baseline within $\sim10\degr$ of the poles, the relative depth is not a monotonic function of ecliptic latitude. 

At 16.5 months, the LVFit PAH map is already deeper than WISE at the ecliptic poles, within 15--35\% of the WISE depth in the other low-latitude fields (Corona Australis and Rosette, $|\beta| < 20\degr$), and 1.4--2 times noisier in the mid-latitude fields (M31, M33, and the Heart and Soul nebulae). Since the noise in these faint regions scales as $N_{\rm pts}^{-1/2}$, we project the depth by the end of the two-year prime survey \citep{Bock_2026_ApJ} and after a two-year extension, based on the number of survey visits obtained so far versus the number expected. By the end of the prime survey, we estimate that the low-latitude fields will achieve parity with WISE or better (noise 0.8--1.0$\times$ that of WISE), while the mid-latitude fields will remain 1.2--1.7 times noisier. A two-year extension would bring most fields to parity (e.g. the Heart and Soul nebulae) or within a few percent of it (e.g. M33), but the mid-latitude fields with the fewest visits (e.g. M31) would remain $\sim25\%$ noisier than WISE, because the current survey strategy places a quarter of all exposures within $15\degr$ of the ecliptic poles. Spreading the exposures of an extension more evenly over the sky would raise the exposure count elsewhere by up to 35\% (a 16\% gain in depth), which would bring even those fields to within a few percent of WISE (e.g. M31 to within $\sim6\%$) by the end of the extension.

\subsection{Reasons for Banding} \label{sec:reasBand}

Working with LVF data presents a challenge for classical methods of photometric and spectroscopic analysis, but also an opportunity. Most methods that attempt to treat LVF data like classical photometry result in banding in the scanning direction, as shown in the top left of Figure~\ref{fig:rosette_methodcompare_full}. More generally, any mismodeling of LVF data will result in banding in the resultant images and so banding can be viewed as a clear indicator of model misspecification. Designing models that naturally do not produce unphysical banding in images resulting from LVF data reductions is necessary, but not sufficient, to have a well-specified model that describes the data.

Because more missions with LVF instruments are coming online \citep[e.g. the PRobe far-Infrared Mission for Astrophysics, PRIMA;][]{Glenn_2025_JATIS}, it is important we codify some methodological lessons from developing SPHEREx reductions for these future surveys. In our experience, there are three main causes of banding: mismodeling either the instrument, a component you care about, or a component you do not care about. In the first case, the simplest example of this comes from roughly approximating the spatially varying spectral response of the detector, for example by defining discrete spectral channels. This is why we are careful to linearly interpolate the spectral response at the highest resolution possible. This means careful laboratory and on-sky calibration are critical for LVF instruments. 

The second case is the one that most astronomers are trained to consider, for example by describing a PAH complex of interest by increasingly complicated substructure. We find that data-driven methods are critical in this case because we aim to discover unknown unknowns about the features of interest from the data. Being data-driven also allows us to defer questions of absolute spectral shape by making inferences only in the spectral-response-convolved space, which is the one that is actually observed.

The third case is the most nefarious, because it results from a component that we might rather just subtract away and forget about. Unlike classical single-band photometry where you might have to subtract off a constant background or a low-order polynomial background to handle a nuisance component, in LVF data, one has to understand both the spatial and spectral properties of the nuisance component in detail.

In all of these cases, to ignore banding is to ignore a misspecified model that for LVF data results not only in a bias, but a spatially varying bias. These biases can be so large as to obscure even the spatial structure of the features of interest (see Section~\ref{sec:compare2others}), but even when significantly suppressed may also impart more subtle biases that impact studies of the diffuse emission correlation function, such as for quantitative studies of interstellar turbulence \citep[e.g.,][]{Miville-Deschenes_2016_AA, Vigoureux:2026}. As a result, it is essential to improve our forward models on all of the above axes. Additionally, it should be an extremely high priority for the community to be producing and consuming estimates of the correlated uncertainties that account for any residual banding.

\subsection{Comparison to Literature} \label{sec:compare2others}

There have been several previous works that make PAH maps from SPHEREx and we briefly compare LVFit with those works.

\begin{itemize}
    \item \citet{Cukierman_2026_ApJS} produce maps up to the native $6\farcs15$ pixel scale of SPHEREx \citep{Hui_2026_ApJS, SPHEREx_ExpSupp_2026} for many test fields using a channel-binning approach, which is improved by interpolation between the bins to reduce banding. 
    \item \citet{Murgia_2026_ApJ} follow the method of \citet{Cukierman_2026_ApJS}, but smooth to $20\arcmin$ FWHM to reduce remaining banding effects and study large scale correlations.
    \item \citet{Hora_2026_ApJ} use channel maps, but apply a ``correction factor'' that weights the channels by a fixed spectral shape assumed for the PAH. While not applied in the forward model direction, this should be equivalent to LVFit with a single, non-data-driven component, though it does not give access to the residuals and $\chi^2$ of the fit.
    \item \citet{Boersma_2026_ApJ} fit analytic continuum and Gaussian models of the PAH peaks integrated over top-hat approximations of the spectral response at $7\farcm2$ apertures on a $7\times7$ grid.
    \item \citet{Li_2026_arXiv} make channel maps of the LMC/SMC at $2\arcmin$.
    \item \citet{Cepeda-Arroita_2026_arXiv} forward model \healpix $N_{\rm side} = 2^{12}$ ($\sim0\farcm9$) pixels with a $2\farcm6$ FWHM beam, using analytic Drude, Gaussian, and continuum models with a nonlinear fit to channel-binned measurements using per-channel spectral response functions.
\end{itemize}

\begin{figure*}[htb]
    \centering
    \includegraphics[width=\linewidth]{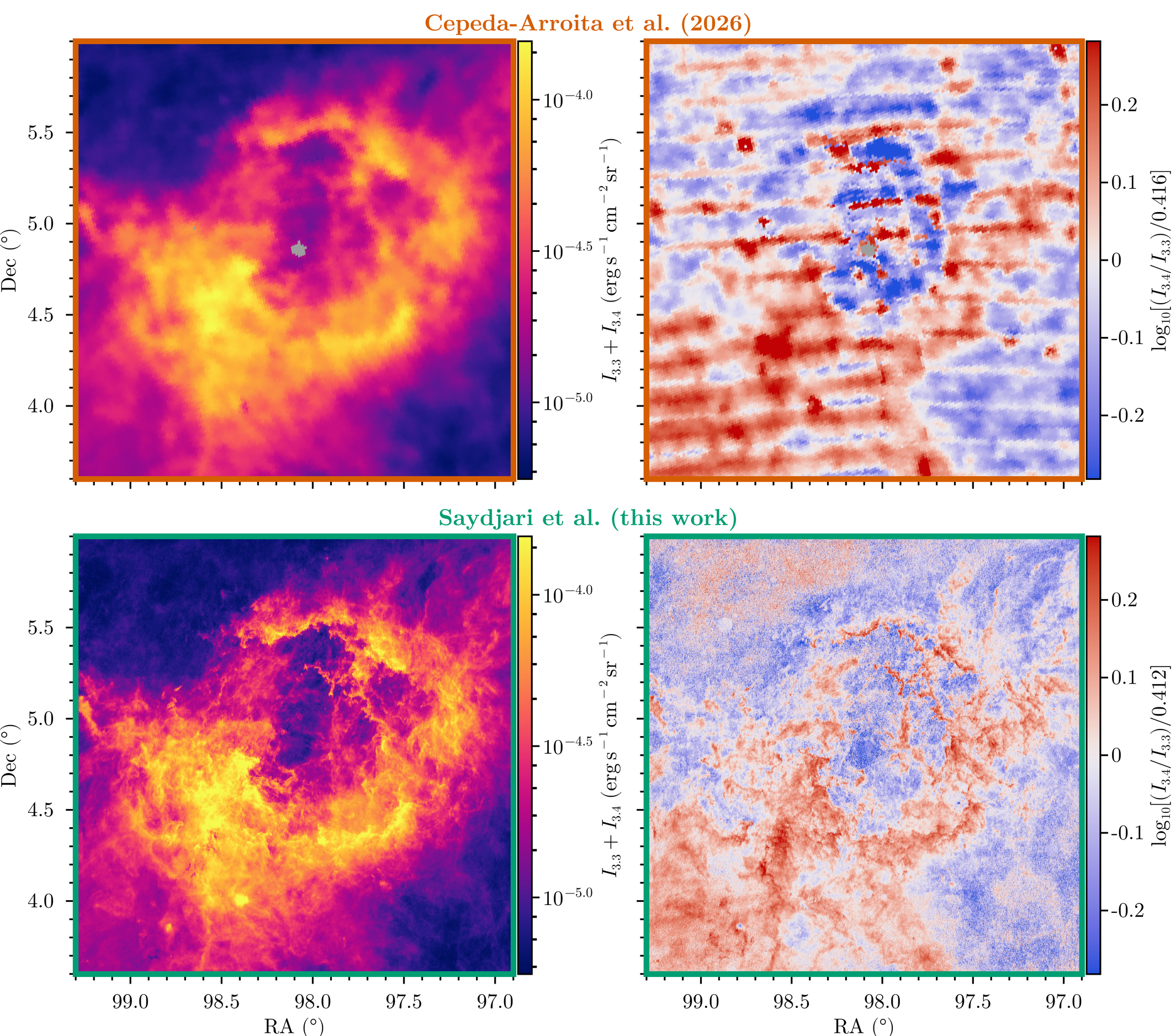}
    \caption{Comparison of the \citet{Cepeda-Arroita_2026_arXiv} SPHEREx PAH maps (top) with the LVFit maps produced in this work (bottom) for the Rosette nebula. Left: Summed $3.3+3.4$\um line surface brightness on a common logarithmic color scale. Right: $\log_{10}[(I_{3.4}/I_{3.3})/\mathrm{median}]$, each normalized by its own median over the panel (0.416 and 0.412), on a common $\pm0.28$\,dex scale. The \citet{Cepeda-Arroita_2026_arXiv} maps are their unfiltered products at native resolution (\healpix $N_{\rm side} = 2^{12}$, $2\farcm6$ FWHM beam) and the LVFit maps are shown at native resolution ($N_{\rm side} = 2^{15}$ pixels, $8\farcs5$ FWHM beam).}
    \label{fig:rosette_cepeda}
\end{figure*}

The main differences in this work are that we forward model the data by the highest spatial resolution representation of variation in the spectral response of the detector, maintain the full spatial resolution of SPHEREx, and use PAH lineshape models that are initialized from observations at higher spectral resolution with learned data-driven corrections. The combination of these choices naturally produces images without apparent banding. 

We compare the LVFit and \citet{Cepeda-Arroita_2026_arXiv} maps of the Rosette nebula in Figure~\ref{fig:rosette_cepeda}, because their approach is the most similar to our own. On large scales ($2\farcm6$) the maps agree to $\sim4\%$ for the summed $3.3+3.4$\um surface brightness and $\sim1\%$ for the $3.4/3.3$\um ratio. However, residual banding along the SPHEREx scan direction is visible both subtly in the PAH surface brightness map and clearly in the $3.4/3.3$\um ratio map in the \citet{Cepeda-Arroita_2026_arXiv} map (Figure~\ref{fig:rosette_cepeda}, top row). The banding is not visible in either the surface brightness or ratio map using LVFit (Figure~\ref{fig:rosette_cepeda}, bottom row). While we only show the Rosette nebula, this general trend of LVFit suppressing banding in both the surface brightness and ratio maps holds for all of the regions we have examined.

\section{Conclusions} \label{sec:conc}

In this work, we have demonstrated LVFit, a linear, data-driven, forward-modeling approach to SPHEREx LVF data that provides access to maps of diffuse emission close to the native pixel scale of SPHEREx ($6\farcs4$ pixels, $8\farcs5$ FWHM angular resolution). This approach allowed us to cleanly separate overlapping components of continuum, 3.3\um PAH, 3.4\um PAH, and hydrogen recombination line emission, without the significant banding artifacts that plague other literature SPHEREx reductions. We applied our method to make maps of a wide variety of testbeds that stress test angular resolution, spectral complexity, sensitivity, and spatial deblending. We validate our results against external datasets tracing PAHs, ionized gas, and neutral H gas from WISE, Spitzer, MDW, and LGLBS, highlighting synergies with those datasets as well as upcoming data releases from the SDSS-V LVM. While we only show maps of 3.3\um PAH, 3.4\um PAH, and Paschen-$\alpha$ emission for specific regions, the methodology presented here can be straightforwardly extended to other emission features in the SPHEREx wavelength range and is scalable to the entire sky (in $\sim 60$k core-h).

However, there are several limitations of the methodology in this work that impact the quality of the final map. Future work would benefit from the following methodological advances.

\paragraph{Model PSF}

Here we model the SPHEREx data at the pixel level, ignoring any correlations between measurements introduced by the point-spread function. This significantly impacts our ability to model point sources (see below) and limits our ability to resolve substructures in the diffuse emission smaller than the PSF scale, which is in principle possible via a forward model of different exposures with a variety of subpixel shifts. Furthermore, there are real variations in the point-spread function over the detector plane. Future work could explicitly learn and forward model the spatially varying point spread function to recover more and higher-resolution PAH emission with fewer star artifacts as a result.

\paragraph{Model point sources}

We do not explicitly model point sources. We mask and conditionally infill over them. By jointly forward modeling the diffuse emission and the point sources, we expect to better recover structures in the diffuse emission that are coincident with point sources compared to our only statistically accurate infills. Improved PSF models (see above) and wider wavelength coverage should facilitate better point source ``removal.''

\paragraph{Fully data-driven basis set}

Here we take a practical approach to determining a fixed set of basis vectors. We use two PAH basis vectors learned as data-driven perturbations from high-resolution JWST spectra, three continuum components normalized to unity at 3.3\um (constant, $\lambda^2$, and $\lambda^{-2}$), and one theoretically generated basis vector for hydrogen recombination lines in regions with strong ionized gas emission.

Given the large number of SPHEREx measurements available, it should be possible to learn the line shapes directly from the data, including additional types of line variation and components that are not anticipated in advance. These could potentially be estimated in a preprocessing step but would ideally be learned jointly with the maps, reducing the possibility of model misspecification. One difficulty we have run into, when adding ro-vibrational $\rm{H}_2$ emission line basis vectors, is that while their amplitudes may be easy to fit in regions of high signal, it may be necessary to regularize the amplitude of rarer basis vectors so that they do not bias measurements of the other components in regions of low signal-to-noise. This regularization fits naturally into the variational inference approach we describe below.

\paragraph{Data-driven exposure corrections}

There are several poorly constrained components in SPHEREx observations that are expected to vary at the exposure level, which are difficult to measure and correct when working at the pixel level and might require alternating passes between pixel-level and exposure-level fit parameters to resolve. The two most obvious examples are ``shuttle glow,'' emission from the spacecraft interacting with atomic oxygen in the atmosphere that is strongest when the pointing has a small angle relative to the direction of motion \citep{SPHEREx_ExpSupp_2026}, and zodiacal light from scattered sunlight and thermal emission of interplanetary dust.

For zodiacal light, we rely on a model provided by the SPHEREx collaboration based on \citet{Kelsall_1998_ApJ} to perform a point-estimate subtraction of zodiacal light from each image before fitting pixel values. However, we also expect deviations from the model which cannot be fully captured by static-sky continuum subtraction as they are determined by orientation of the telescope pointing relative to solar system objects (e.g. Jupiter, Mars) based on the Earth's orbital phase and are not the same in all exposures which point to a given location of the sky. Failure to properly model the zodiacal light is one form of model misspecification that we suspect makes banding at low ecliptic latitudes so difficult to eliminate, because it is where the zodiacal light is strongest. To solve this problem, we are pursuing data-driven models of the zodiacal light or shuttle glow that are constrained to depend only on the differences between exposures that we expect to be physically and causally linked to those signals (e.g., angle relative to spacecraft motion).

\paragraph{Time dependence}
We are modeling the astrophysical sky as if it were perfectly static. However, as we discuss above, moving solar system objects violate that assumption and had to be masked. Similarly, parallax and proper motion of point sources cannot be modeled in LVFit. Brightness variations from transient events (e.g., supernovae) or periodic variable stars are also outside of our framework, though they are masked by our current approach that disregards point sources. However, any approach that attempts to model the point sources will have to contend with variable stars. For example, our preliminary efforts in this direction have struggled to use a static model to describe a bright Mira variable in our Rosette nebula footprint. In addition, diffuse structures may have observable proper motions if they are very nearby \citep[e.g.,][]{Piecka_2025_A&A} or brightness variations \citep[e.g., supernova light echoes;][]{Dwek_2008_ApJ}, which, if left unmodeled, could decrease the quality of our model fit to the data.

\paragraph{Spatial correlations}
Our work treats data-constrained model pixels independently, averaging over a small Gaussian beam. This procedure does not acknowledge the actual power-law spectrum of the diffuse emission, which exhibits more correlation at larger scales and less correlation at scales below the beam size. Proper treatment of these correlations could produce maps with simultaneously less noise, fewer artifacts, higher resolution, and improved separation of diffuse emission into components which differ in their spatial structure. One promising option is to use a Gaussian process prior with an appropriate covariance kernel for the entire sky, similar to our approach for the small number of unconstrained masked pixels in this work. One challenge with this approach is representing such a prior over large areas of sky at high resolution, but this is feasible with recently developed methods such as GraphGP \citep{Dodge_2026_arXiv}.

\paragraph{Uncertainty quantification}

Our method is a simple conditional (on the basis vectors we learned) linear optimization and the formal uncertainties on the fit are only close to the truth in the faint-sky limit. We had to empirically calibrate our reported uncertainties in Section~\ref{sec:sens} for the bright-sky limit to account for model errors and the fact that we ignore correlated uncertainties. A full posterior that accounts for the covariance between map pixels and yields internally consistent and well-calibrated errors would be valuable for differentiating real structure from noise, propagating uncertainty into subsequent measurements made from these maps, and trusting that systematics and the instrument response are completely understood.

\paragraph{Future directions}

Many of the aforementioned limitations can in principle be solved by adding additional hyperparameters to the model (e.g., PSF, basis vectors, stellar spectra, zodi), choosing appropriate priors (e.g., Gaussian processes), and drawing correlated samples from the joint posterior. This problem is no longer linear in its parameters nor can it be solved independently for each of its billions of model pixels, presenting a major challenge for many techniques. Fortunately, algorithms such as Metric Gaussian Variational Inference \citep{Knollmuller_2019_arXiv} and Geometric Variational Inference \citep{Frank_2021_Entrp} have been developed to solve such Bayesian inverse problems at scale with acceptable computational cost. These methods have been applied successfully to 4D maps with three spatial dimensions and one scalar value previously (Frank \& Saydjari et al. 2026, in prep.), and we are working on a joint spatial-spectral application to SPHEREx (Dodge et al. 2026, in prep.).

\section{Data Availability} \label{sec:dataavil}

We release the maps of every field in this paper in two forms. The primary products are HDF5 tables on the \healpix $N_{\rm side} = 2^{15}$ grid. For convenience, we also provide FITS rasters of the main maps on a $3\farcs22$ grid in the frame of each field. The FITS files hold one image extension for each primary map product. These are the infilled 3.3\um PAH, 3.4\um PAH, and hydrogen recombination line maps, the latter only on fields where it was fit (Section~\ref{sec:infill}). These maps include both the mean and two conditional draws of the infill along with their calibrated uncertainties. The files also include the continuum surface brightness at 3.29\um, the reduced $\chi^2$, the number of measurements per pixel, and the mask. The HDF5 tables carry additional fields that completely describe the fit. They give the amplitude of every basis vector and its formal uncertainty, the fit line values before infilling, the pixel coordinates, and the list of exposures that contributed to each field. For masked pixels, we include a bitmask indicating why they were masked (Section~\ref{sec:mask}): bit 0 for insufficient coverage or a failed fit, bit 1 for a star disk, bit 2 for a residual outlier, bit 3 for a residual outlier near a star, bit 4 for the ratio-based growth of a bright star's disk, and bit 5 for pixels added by the final closure step. 

All of these data products, the PAH basis vectors, the spectral response model, our list of detector pixels with anomalous residuals (Section~\ref{sec:linfit}), the third-party maps used in our comparisons, and the code necessary to produce the maps from the SPHEREx images and every figure and number in this paper from the maps are provided in a Zenodo repository at \dataset[doi:10.5281/zenodo.24068379]{https://doi.org/10.5281/zenodo.24068379}. The SPHEREx images themselves are not redistributed, but the archive includes the measurement-level inputs of the two figures made from individual measurements (Figures~\ref{fig:pixel_schematic} and \ref{fig:moon_edge}). The maps were made from the Quick Release 2 and 3 spectral images, which are available from the NASA/IPAC Infrared Science Archive (IRSA)\footnote{\url{https://irsa.ipac.caltech.edu/Missions/spherex.html}} \citep{SPHEREx_ExpSupp_2026}. The total Zenodo archive is $\sim$35 GB unzipped. The figures in this article are only provided at 300 dpi, but the full-resolution versions of every figure are also included in the repository.

\begin{acknowledgments}
A.K.S. acknowledges that support for this work was provided by NASA through the NASA Hubble Fellowship grant HST-HF2-51564.001-A awarded by the Space Telescope Science Institute, which is operated by the Association of Universities for Research in Astronomy, Inc., for NASA, under contract NAS5-26555. C.H. is supported by the National Science Foundation Graduate Research Fellowship under Grant Number DGE-2234660. B.D. is supported by the National Science Foundation Graduate Research Fellowship under grant number DGE-2146755. S.E.C. acknowledges support from the U.S. National Science Foundation under NSF CAREER grant No. AST-2441452, and from an Alfred P. Sloan Research Fellowship. The Flatiron Institute is a division of the Simons Foundation.

We acknowledge helpful discussions with
  Matt Ashby,
  Chris Clark,
  Kinsey Cronin,
  Julianne Dalcanton,
  Bruce Draine,
  Doug Finkbeiner,
  Gregory Green,
  Tom Hilder,
  Joe Hora,
  Howard Hui,
  Adam Leroy,
  Claire Murray,
  Hans-Walter Rix,
  Karin Sandstrom,
  Eddie Schlafly, and
  Liz Tarantino.
A.K.S. acknowledges Sophia S\'{a}nchez-Maes for helpful discussions and much support. This work was supported in part by the Carnegie Theoretical Astrophysics Center (CTAC) through its Theory Scholar program. The authors acknowledge the Interstellar Institute and the Institut Pascal at Université Paris-Saclay for hosting discussions that nourished the development of the ideas behind this work. This work resulted from collaborations supported by the Kavli Institute for Particle Astrophysics and Cosmology through a Workshop Program grant to Philipp Frank for the Mapping the Diffuse Milky Way workshop held at Stanford on October 20–24, 2025, made possible by an endowment by the Kavli Foundation.

Above all, it is a pleasure to thank the NASA SPHEREx team for generously making these data public on a fast timescale.

\end{acknowledgments}

\begin{contribution}
A.K.S. co-conceived of the project, developed the methods, wrote and ran the code, produced the figures, co-ideated testbeds, and co-wrote the first draft of the manuscript. C.H. developed the database key to the speed of our method, managed data ingestion, and provided feedback on the manuscript. B.D. co-conceived of the project, contributed to understanding banding, tested alternative methods, and wrote parts of the manuscript. P.F. helped develop alterative methods to overcome limitations of LVFit and provided feedback over the course of the project. D.W.H. co-conceived of the project, provided feedback during method development, and wrote parts of the manuscript. S.E.C. co-ideated testbeds, provided feedback over the course of the project, and co-wrote the first draft of the manuscript. All authors have read and agree to the final manuscript.
\end{contribution}

\section*{Generative AI Disclosure} The original concept, methods, code base, and first draft of the manuscript were written by the authors. We used a Claude Code agent (Anthropic; Fable 5.1 and Opus 5.5) to refactor and automate the code base, execute and monitor some of the batch fits and calibration tests, polish and write some figure-production scripts, draft and edit some passages of the text at our direction, audit some of the methods description and numbers against the code and data products, and locate and format some bibliography entries. All code, text, figures, and references were reviewed by the authors, and all text was edited by them. No AI tool generated data, images, or scientific results beyond running the authors' pipeline, which is fully available as described in Section~\ref{sec:dataavil}. The authors take full responsibility for the content of this work.

\facilities{SPHEREx, IRSA, WISE, Spitzer (IRAC), VLA, GBT, FLWO:2MASS, CTIO:2MASS}

\software{Julia \citep{Bezanson_2017_SIAMRev}; Makie.jl \citep{Danisch_2021_JOSS}; Healpix.jl \citep{Tomasi_2021_ASCL} and \healpix \citep{Gorski_2005_ApJ}; DataFrames.jl \citep{BouchetValat_2023_JSS}; DuckDB \citep{Raasveldt_2019_SIGMOD} and Apache Parquet \citep{Apache_Parquet}; HDF5.jl with HDF5 \citep{HDFGroup_2025_Zenodo}; JLD2.jl; FITSIO.jl with cfitsio \citep{Pence_2010_ASCL}; WCS.jl with WCSLIB \citep{Greisen_2002_AA, Calabretta_2002_AA}; SkyCoords.jl; SPICE.jl with the SPICE toolkit \citep{Acton_1996_PSS, Acton_2018_PSS}; StatsBase.jl \citep{Lin_2026_Zenodo}; Interpolations.jl \citep{Kittisopikul_2022_Zenodo}; LsqFit.jl; CSV.jl \citep{Quinn_2026a_Zenodo}; JSON3.jl \citep{Quinn_2025_Zenodo}; HTTP.jl \citep{Quinn_2026b_Zenodo}; CodecZlib.jl; StructArrays.jl; LaTeXStrings.jl; ColorSchemes.jl with colorcet colour maps \citep{Kovesi_2015_arXiv}; NumPy \citep{Harris_2020_Nature}; PyNeb \citep{Luridiana_2015_AA}}

\bibliography{PAH}{}
\bibliographystyle{aasjournalv7}

\end{document}